\documentclass[%
 aip,
 amsmath,amssymb,
 reprint,%
]{revtex4-1}

\usepackage{graphicx}
\usepackage{dcolumn}
\usepackage{bm}
\usepackage{lineno}

\usepackage[utf8]{inputenc}
\usepackage[T1]{fontenc}
\usepackage{mathptmx}

\begin{document}

\preprint{AIP/123-QED}

\title[Tutorial]{Linking information theory and thermodynamics to spatial resolution in photothermal and photoacoustic imaging}

\author{P. Burgholzer}
 \affiliation{Research Center for Non Destructive Testing (RECENDT), 4040 Linz, Austria}
  \email{peter.burgholzer@recendt.at}

\author{G. Mayr}
\affiliation{ 
Josef Ressel Center for Thermal NDE of Composites, University of Applied Sciences Upper Austria, 4600 Wels, Austria}

 \author{G. Thummerer}
\affiliation{ 
Josef Ressel Center for Thermal NDE of Composites, University of Applied Sciences Upper Austria, 4600 Wels, Austria}

\author{M. Haltmeier }%
\affiliation{Department of Mathematics, University of Innsbruck, 6020 Innsbruck, Austria}

\date{\today}

\begin{abstract}
In this tutorial, we combine the different scientific fields of information theory, thermodynamics, regularization theory and non-destructive imaging, especially for photoacoustic and photothermal imaging. The goal is to get a better understanding of how information gaining for subsurface imaging works and how the spatial resolution limit can be overcome by using additional information. Here, the resolution limit in photoacoustic and photothermal imaging is derived from the irreversibility of attenuation of the pressure wave and of heat diffusion during propagation of the signals from the imaged subsurface structures to the sample surface, respectively. The acoustic or temperature signals are converted into so-called virtual waves, which are their reversible counterparts and which can be used for image reconstruction by well-known ultrasound reconstruction methods. The conversion into virtual waves is an ill-posed inverse problem which needs regularization. The reason for that is the information loss during signal propagation to the sample surface, which turns out to be equal to the entropy production. As the entropy production from acoustic attenuation is usually small compared to the entropy production from heat diffusion, the spatial resolution in acoustic imaging is higher than in thermal imaging. Therefore, it is especially necessary to overcome this resolution limit for thermographic imaging by using additional information. Incorporating sparsity and non-negativity in iterative regularization methods gives a significant resolution enhancement, which was experimentally demonstrated by one-dimensional imaging of thin layers with varying depth or by three-dimensional imaging, either from a single detection plane or from three perpendicular detection planes on the surface of a sample cube.  
\end{abstract}

\maketitle

%

\section{\label{sec:Introduction}Introduction}

Metrology can be described as collecting information from samples. Imaging techniques provide a lot of information, as is indicated by the saying ``a picture is worth a thousand words''. Non-destructive evaluation (NDE) or biomedical imaging often image deep structures in the samples interior, without destroying the samples. Therefore, an information theoretical viewpoint is instructive to determine the amount of information on subsurface or other embedded structures, which can be gained by measurements on the sample surface. Information theory is a branch of applied mathematics, electrical engineering, and computer science involving the quantification of information. Information theory was developed by Shannon \cite{Shannon.1948} to find fundamental limits on signal processing operations such as compressing data and on reliably storing and communicating data. Information processing is a physical activity, which has to obey the laws of (non–equilibrium) thermodynamics. This was originally recognized by Szilárd \cite{Szilard.1929}, and Landauer\cite{Landauer.1991} captured it with his aphorism: "Information is physical".\par
\begin{figure}[h!]
\includegraphics[trim={10.0cm 0cm 12.0cm 0cm},clip,width=\linewidth]{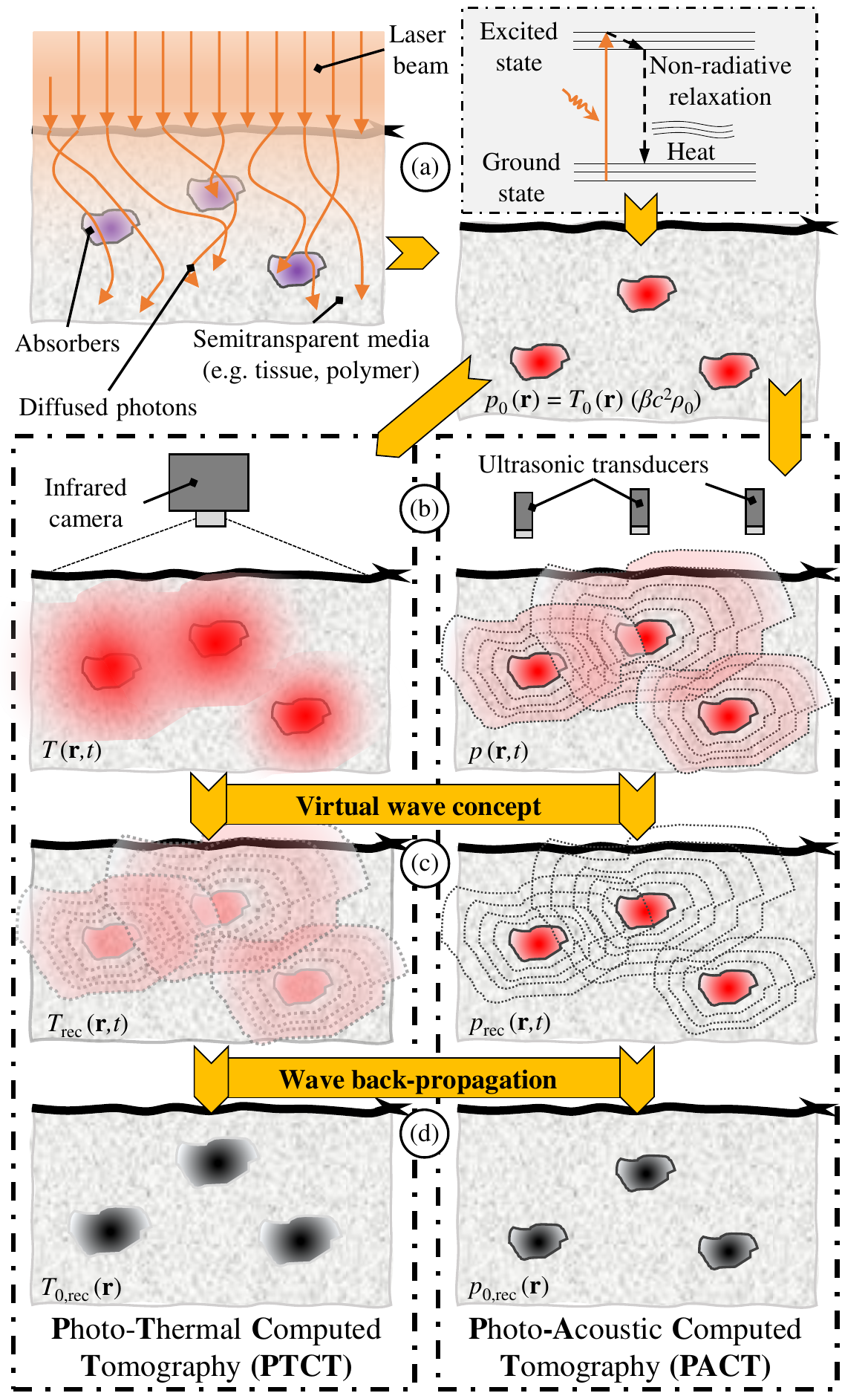}
\caption{\label{fig:Fig_Imaging_structures} Schematic sketch of the imaging of absorbers within a sample inspired by Li et al.\cite{Li.2018}: (a) Sample containing absorbers as an internal structure, which should be imaged by suddenly heating this structure, e.g. by absorbing scattered light from a laser pulse. (b) Propagation of induced signals, such as thermal or pressure waves to the sample surface, where they are detected. Signals are blurred due to heat diffusion or acoustic attenuation, respectively. (c) This signal blurring, which reflects noise and entropy production (section \ref{sec:thermodynamic}), can be compensated to a certain extent when calculating the virtual wave. The virtual wave is a solution of the ideal wave equation without diffusion or attenuation. Calculating the virtual wave is an ill-posed inverse problem that needs regularization (Eq. (\ref{Eq:linear_relation})). (d) Image reconstruction by wave back-propagation, either of the virtual waves gained from the temperature signals in Photo-Thermal Computed Tomography (PTCT) or of the attenuation compensated virtual acoustic waves in Photo-Acoustic Computed Tomography (PACT).}
\end{figure}

Imaging of subsurface features from data measured on the sample surface usually results in an ill-posed or an ill-conditioned inverse problem, which needs regularization \cite{Hansen.1998,Aster.2018,Scherzer.2009}, as illustrated in Fig. \ref{fig:Fig_Imaging_structures} for photoacoustic and photothermal imaging. Regularization typically involves additional assumptions, such as the smoothness of the solution. Usually, scientists working on the inverse problem of subsurface imaging are barely working on non-equilibrium thermodynamics and information theory, and vice versa. This tutorial should enable researchers from both sides to get more insight from the counterpart. The models used for subsurface imaging and for thermodynamics are kept simple in the beginning, allowing to investigate the essential connections between thermodynamics and regularization of inverse problems. As a first step, a one-dimensional (1D) model without boundaries and a Dirac delta-like structure at a varying depth is used for imaging, either inducing thermal waves described by heat diffusion or acoustic pressure signals showing acoustic attenuation as in a liquid. For describing the relevant results from non-equilibrium thermodynamics, we follow a paper from Esposito and Van den Broeck about the 
``Second law and Landauer principle far from equilibrium'' \cite{Esposito.2011}, where we describe the photothermal or photoacoustic measurement process in terms of thermodynamics. The description can be simplified as the Hamiltonian does not change in time. A short laser light pulse (Dirac delta-like) is used for excitation, which brings the sample suddenly out of thermal equilibrium by heating the structures via optical absorption and thereby generating an initial pressure distribution, and then the system returns slowly to equilibrium either by heat diffusion (
``thermal'') or by attenuation of the pressure wave (``acoustic'').\par

After presenting these simple models for the wave propagation and the thermodynamic approach, they are linked in frequency domain to evaluate the physical background of regularization. Later on, more realistic samples in higher dimensions and with realistic boundary conditions are described. But before stepping into the details, in the remaining part of the Introduction section previous work in subsurface imaging and in thermodynamics is described.\par
In former times, images always have shown what could be detected with human eyes. Looking under the sample surface in a non-destructive way was possible only for optically transparent samples. This situation changed drastically with the discovery of X-rays by  Röntgen in 1895. Nowadays, the interaction of sample structures with electromagnetic waves in the whole frequency range from eddy current (EC), radar, terahertz (THz), infrared (IR), up to UV, and X-ray radiation is used for imaging, but also with particles or elastic (acoustic) waves. The amount of information, which can be gained by measurements on the sample surface, depends on the interaction of these waves or particles with structures of the sample, such as scattering or absorption, and on the detector (e.g. sensitivity or bandwidth). For interior structures, the available information for imaging is limited also by the information loss during wave propagation from the imaged structures to the sample’s surface, caused by diffusion, dissipation, or scattering (Fig. \ref{fig:Fig_Imaging_structures} or Burgholzer et al. \cite{Burgholzer.2013}). A unifying framework for treating diverse diffusion-related periodic phenomena under the global mathematical label of diffusion-wave fields has been developed by Mandelis \cite{Mandelis.2001}, including thermal waves, charge-carrier-density waves, diffuse-photon-density waves, but also modulated eddy currents, neutron waves, or harmonic mass-transport diffusion waves.\par

There have been made several attempts to compensate the diffusion, dissipation, or scattering during wave propagation to get a higher resolution for the reconstructed images of the samples interior. We have shown that thermodynamical fluctuations are the cause for an entropy production, which is equal to the information loss and limits this compensation\cite{Burgholzer.2013}. Therefore, also the spatial resolution for non-destructive imaging (NDI) at a certain depth is limited. For waves that satisfy the wave equation, no information is lost during the propagation, because the time reversed wave is a solution of the same wave equation. The inverse problem of 
``back-projection'' can be exactly solved by, e.g. time reversal \cite{Burgholzer.2007}. It is called a virtual wave signal, because in reality always dissipation, diffusion, or scattering causes a loss of information during propagation. For attenuated acoustic waves \cite{Szabo.2014, LaRiviere.October23292005, LaRiviere.2006, Ammari.2012, DeanBen.2011, Kowar.2012, Burgholzer.2007b, Burgholzer.2010, Burgholzer.2010b, Burgholzer.2011, Treeby.2010, Treeby.2010b, Treeby.2013} and thermal diffusion \cite{Burgholzer.2013, Burgholzer.2017, Burgholzer.2015, Burgholzer.2018, Burgholzer.2017b, Gershenson.2018, Gershenson.2019} a linear relation between the time discretized measured signal $\mathbf{Y}_\text{meas}$ and the virtual wave signal $\mathbf{Y}_\text{virt}$, which is a solution of the wave equation, was established, and can be written as
\begin{equation}
    \mathbf{Y}_\text{meas} = \mathbf{M} \, \mathbf{Y}_\text{virt}.
    \label{Eq:linear_relation}
\end{equation}
The matrix $\mathbf{M}$ depends on the kind of wave propagation and was calculated already for acoustic waves having a power law attenuation and for heat diffusion. The solution of Eq. (\ref{Eq:linear_relation}) constitutes an ill-posed or ill-conditioned inverse problem. The direct inversion of this equation would therefore cause severe noise amplification resulting in useless solutions. Therefore, regularization methods are used for the inversion. This is a consequence of the information loss during signal propagation to the surface. $\mathbf{Y}_\text{meas}$ could be the measured acoustic pressure or temperature, and $\mathbf{Y}_\text{virt}$ could be the virtual pressure or temperature, respectively.\par
Eq. (\ref{Eq:linear_relation}) was derived for attenuated acoustic waves and thermal diffusion, but the concept of deriving the resolution limit from entropy production seems to be useful for non-destructive and biomedical imaging in general. For example, applications of a transformation of the diffusive electromagnetic wave into a wave field were shown by Lee et al. \cite{Lee.1989, Lee.1993} and Gershenson \cite{Gershenson.1997} for geophysical inverse problems.  The physical reason for the entropy production are fluctuations, which is the noise added to an ideally noise-free signal and can be statistically described by stochastic processes \cite{Burgholzer.2013}. The ratio between the ideal signal amplitude and the noise amplitude is called the signal-to-noise ratio (SNR) and plays an essential role in this tutorial. Noise is identified as the thermodynamic fluctuations around a certain time varying mean-value and is shown in section \ref{sec:thermodynamic} to be the mechanism for entropy production and information loss.
\par
One comprehensive letter about the relevant non-equilibrium thermodynamics, the second law and the connection between entropy production and information loss, was published in 2011 by Esposito and van den Broeck \cite{Esposito.2011}. They showed that for two different non-equilibrium states evolving to the same equilibrium state, the entropy production $\Delta_i S$ during the evolution from one state to the other is equal to the information loss $\Delta I = k_\text{B} \Delta D$, where $k_\text{B}$ is the Boltzmann constant and $\Delta D$ is the difference of the Kullback-Leibler divergence $D$, also called relative entropy of these states. $D$ is a measure of how “far” a certain state is away from equilibrium \cite{Cover.2006}. The entropy production for the macroscopic states with small fluctuations around equilibrium turns out to be, in a good approximation, equal to the dissipated energy $\Delta Q$, which is the dissipated heat or the heat transported by diffusion, divided by the mean temperature $T_\text{mean}$, so $\Delta_i S = \Delta Q / T_\text{mean} = k_\text{B} \Delta D$ \cite{Burgholzer.2015}.\par
The thermodynamic fluctuations in macroscopic samples are usually so small that they can be neglected – but not for ill-posed inverse problems such as described by the inversion of Eq. (\ref{Eq:linear_relation}). The fluctuations are highly amplified due to the ill-posed problem of image reconstruction. As long as the macroscopic mean-value-equations describe the mean entropy production, the resolution limit depends only on the amplitude of the fluctuations and not on the actual stochastic process including all correlations \cite{Burgholzer.2015}. Therefore, we have used in the past a simple Gauss-Markov process to describe the measured signal as a time-dependent random variable \cite{Burgholzer.2013}. In addition, we have performed preliminary work to describe heat diffusion as a Wiener process and have compared for a simulated signal the spatial resolution limits derived in temporal and spatial frequency domain, $\omega$- and $k$-space, respectively \cite{Burgholzer.212015}. Experimentally, these theoretical resolution limits were verified by thermographic reconstructions of inclined steel rods in an epoxy sample, heated by a short light pulse \cite{Burgholzer.2018} (section \ref{sec:single_detector_plane}).

\section{\label{sec:Imaging_model}Photoacoustic and photothermal imaging model}
\label{sec:Photothermal_acoustic_model}

In photothermal and photoacoustic imaging a short light pulse, e.g. from a laser, is used to heat subsurface structures by the absorbed (scattered) light, such as blood vessels in tissue or light absorbing structures in epoxy resin, creating a temperature distribution $T_0 (\mathbf{r})$ at location $\mathbf{r}$ immediately  after the pulse (Fig.\ref{fig:Fig_Imaging_structures}a). This sudden temperature increase causes an initial pressure distribution $p_0 (\mathbf{r})$ proportional to the absorbed optical energy density $A(\mathbf{r}) = T_0 (\mathbf{r}) C_p \rho_0$, with a dimensionless material constant $\Gamma = \beta c^2 / C_p$, the Grüneisen coefficient \cite{Gusev.1993, Kostli.2001, Xu.2002}. The initial pressure distribution can be written as
\begin{equation}
    p_0(\mathbf{r}) = \Gamma A (\mathbf{r}) = \frac{\beta c^2}{C_p} A (\mathbf{r}) = \beta c^2 \rho_0 T_0 (\mathbf{r}),
    \label{Eq:Initial_pressure}
\end{equation}
\noindent
where $C_p$ is the specific heat capacity, $\rho_0$ is the ambient density, $c$ is the sound speed, and $\beta$ is the thermal volume expansion coefficient at constant pressure. The initial pressure distribution causes a virtual pressure wave $p_\text{virt} (\mathbf{r},t)$ as a function of space $\mathbf{r}$ and time $t$. This wave is called ``ideal'' as no acoustic attenuation or dispersion during propagation is assumed and is ``virtual'' as attenuation and dispersion are always present for real waves, even if acoustic attenuation can often be neglected for lower frequencies. The wave equation describes the acoustic pressure $p_\text{virt} (\mathbf{r},t)$ as a function of space $\mathbf{r}$ and time $t$ and can be written as \cite{Kostli.2001, Xu.2002}
\begin{equation}
    \left(\nabla^2 - \frac{1}{c^2} \frac{\partial^2}{\partial t^2}\right) p_\text{virt}(\mathbf{r},t) = - \frac{1}{c^2} \frac{\partial}{\partial t} p_0 (\mathbf{r})\delta(t),
    \label{Eq:Wave_equation}
\end{equation}
\noindent
where $\nabla^2$ is the Laplacian (second derivative in space). The source term on the right side of Eq. (\ref{Eq:Wave_equation}) ensures that the pressure just after the short excitation pulse, which is modeled by the temporal Dirac delta function $\delta(t)$, is the initial pressure distribution $p_0 (\mathbf{r})$. The same equation is valid for a virtual temperature wave, which is defined as $T_\mathrm{virt} := p_\mathrm{virt}(\mathbf{r},t) / \left ( \beta c^2 \rho_0 \right)$ and gives
\begin{equation}
    \left(\nabla^2 - \frac{1}{c^2} \frac{\partial^2}{\partial t^2}\right) T_\text{virt}(\mathbf{r},t) = - \frac{1}{c^2} \frac{\partial}{\partial t} T_0 (\mathbf{r})\delta(t)\,.
    \label{Eq:Virt_wave_equation}
\end{equation}
\noindent
The virtual wave is the induced virtual pressure wave, but multiplied by a material constant to get a temperature measured in Kelvin. As it has no direct physical representation but is a mathematical model for thermographic reconstruction, it is called ``virtual temperature wave''.\par
The actual temperature evolution $T(\mathbf{r},t)$ can be described by the heat diffusion equation \cite{Carslaw.19591986printing}
\begin{equation}
    \left(\nabla^2 - \frac{1}{\alpha} \frac{\partial}{\partial t}\right) T(\mathbf{r},t) = - \frac{1}{\alpha} T_0 (\mathbf{r})\delta(t),
    \label{Eq:Diffusion_equation}
\end{equation}
\noindent
where $T(\mathbf{r},t)$ is the temperature as a function of space and time, and $\alpha$ is the thermal diffusivity, which is assumed to be homogeneous in the sample. This description of the thermal diffusion is based on Fourier’s law, which is valid for macroscopic samples, where the propagation distance is much larger than the phonon mean free path \cite{JoseOrdonezMiranda.2015}.\par
We denote by $\tilde{T}(\mathbf{r},\omega)$ the temperature signal in the frequency domain, the $\omega$–space, which is related to $T(\mathbf{r},t)$  by the temporal Fourier transform 
\begin{subequations}
\label{Eq:Fourier_trafo}
\begin{equation}
    \tilde{T}(\mathbf{r},\omega) = \int^\infty_{-\infty} T(\mathbf{r},t) \exp{(i \omega t)} \text{d}t,
\label{subeq:1}
\end{equation}
and its inverse
\begin{eqnarray}
T(\mathbf{r},t) = \frac{1}{2\pi}\int^\infty_{-\infty} \tilde{T}(\mathbf{r},\omega) \exp{(-i \omega t)} \text{d}\omega \,.\label{subeq:Inverse_fourier}
\label{subeq:2}
\end{eqnarray}
\end{subequations}
\noindent
Similarly, $\tilde{p}_\text{virt}(\mathbf{r},\omega)$ and $\tilde{T}_\text{virt}(\mathbf{r},\omega)$ denote the Fourier transforms of the virtual waves $p_\text{virt}(\mathbf{r},t)$ and $T_\text{virt}(\mathbf{r},t)$, respectively. Taking the Fourier transform according Eq. (\ref{Eq:Fourier_trafo}) of Eqs. (\ref{Eq:Wave_equation}), (\ref{Eq:Virt_wave_equation}), and (\ref{Eq:Diffusion_equation}), and using $\delta(t) = \frac{1}{2\pi} \int_{-\infty}^{\infty} \exp{(-i \omega t)} \text{d}\omega$ results in the Helmholtz equations
\begin{eqnarray}
\left( \nabla^2 + k(\omega)^2 \right) \tilde{p}_\text{virt}(\mathbf{r},\omega) = \frac{i \omega}{c^2}p_0(\mathbf{r})\;, \label{Eq:virt_pressure_wave}
\\
\left( \nabla^2 + k(\omega)^2 \right) \tilde{T}_\text{virt}(\mathbf{r},\omega) = \frac{i \omega}{c^2}T_0(\mathbf{r})\;,
\label{Eq:Virtual_wave_frequency}
\\
\left ( \nabla^2 + \sigma(\omega)^2 \right) \tilde{T}(\mathbf{r},\omega) = -\frac{1}{\alpha}T_0(\mathbf{r})\;,
\label{Eq:Temperature_frequency}
\end{eqnarray}
\noindent
with $k(\omega) \equiv \omega / c$ and $\sigma(\omega)^2 \equiv i\omega/\alpha$. For the virtual pressure wave and the virtual temperature wave the wavenumber $k(\omega)$ is real, for the temperature the wavenumber $\sigma(\omega)$ is complex with a real and imaginary part of equal size to describe the diffusive effect.\par
Acoustic attenuation and dispersion can also be described by a complex wavenumber, but usually the imaginary part is much smaller than the real part. As Stokes\cite{Stokes.2009} could already show in 1845 and later on e.g. Shutilov \cite{Shutilov.1988}, fundamental responses of the material system cause a relaxation time between pressure and density changes, which result in an attenuation of a propagating acoustic wave. For liquids the acoustical absorption $\alpha(\omega)$ increases with the square of the angular frequency $\omega$, that $\alpha(\omega)=\alpha_0 \omega^2$, with the material constant $\alpha_0$ taking into account e.g. the viscosity of the liquid. The pressure caused by a monochromatic acoustic plane wave of frequency $\omega$ in a uniform attenuating medium at a point $x$ and at an instant $t$ can be expressed as
\begin{eqnarray}
p_\omega (x,t)&& = \exp{\left(i(k(\omega)x - \omega t \right)} \exp{\left( -\alpha_0 \omega^2 x \right)}\nonumber\\
&& \equiv \exp{\left( i (K(\omega)x-\omega t) \right)},
\label{Eq:pressure_frequency}
\end{eqnarray}
\noindent
where the acoustic attenuation coefficient $\alpha(\omega)$ can be written as the imaginary part of a complex wavenumber $K(\omega)=k(\omega)+i\alpha(\omega)=\omega/c+i\alpha_0 \omega^2$. For the attenuated acoustic wave the Helmholtz equation 
\begin{equation}
    \left(\nabla^2+K(\omega)^2 \right) \tilde{p}(\mathbf{r},\omega)=\frac{i\omega}{c^2}  p_0(\mathbf{r}).
    \label{Eq:Wave_eq_complex}
\end{equation}
is the same as for the virtual pressure wave in Eq. (\ref{Eq:virt_pressure_wave}), but with the complex wavenumber \cite{LaRiviere.October23292005,LaRiviere.2006,Ammari.2012}.
The source term on the right side of Eq. (\ref{Eq:Wave_eq_complex}) is the same as for the virtual pressure wave $\tilde{p}_\text{virt}(r,\omega)$ in Eq. (\ref{Eq:virt_pressure_wave}) and therefore a direct relation can be derived between the attenuated pressure wave and the virtual pressure wave on the same location $\mathbf{r}$ \cite{LaRiviere.October23292005,LaRiviere.2006,Ammari.2012}. Replacing $\omega$ by $cK(\omega)$ in Eq. (\ref{Eq:virt_pressure_wave}) results in
\begin{equation}
    \left(\nabla^2+K(\omega)^2 \right) \tilde{p}_\text{virt}(\mathbf{r},cK(\omega))=\frac{icK(\omega)}{c^2}  p_0(\mathbf{r}),
\end{equation}
\noindent
which is equal to Eq. (\ref{Eq:Wave_eq_complex}), if multiplied by $\omega/cK(\omega)$, and
\begin{equation}
    \tilde{p}(\mathbf{r},\omega)=\frac{\omega}{cK(\omega)}  \tilde{p}_\text{virt}(\mathbf{r},cK(\omega)).
\end{equation}
\noindent
This is the sought relation between the attenuated pressure signal and the virtual pressure signal. The location $\mathbf{r}$ is the same for $\tilde{p}_\text{virt}$ and $\tilde{p}$, and the relation is valid in all dimensions. Transformation back from $\omega$ – space to the time domain with the inverse Fourier transformation (Eq. (\ref{subeq:Inverse_fourier})) results in 
\begin{eqnarray}
    &&p(\mathbf{r},t) = \frac{1}{2\pi} \int_{-\infty}^\infty \frac{\omega}{cK(\omega)} \tilde{p}_\text{virt} (\mathbf{r},cK(\omega)) \exp{(-i\omega t)}\text{d}\omega \nonumber \\     
    && \phantom{X} \text{with} \nonumber \\
    &&\tilde{p}_\text{virt}(\mathbf{r},cK(\omega)) = \int_{-\infty}^\infty p_\text{virt} (\mathbf{r},t') \exp{(icK(\omega)t')}\text{d}t',
\end{eqnarray}
\noindent
which can be written as
\begin{eqnarray}
    p(\mathbf{r},t) &&= \int_{-\infty}^\infty p_\text{virt} (\mathbf{r},t') M_\text{p}(t,t')\text{d}t', \nonumber \\
    && \phantom{X} \text{with} \nonumber \\
    M_\text{p}(t,t')&& \equiv \frac{1}{2\pi} \int_{-\infty}^\infty \frac{\omega}{c K(\omega)} \exp{(i c K (\omega)t')} \exp{(-i\omega t)} \text{d} \omega \nonumber\\
    && \thickapprox \frac{1}{2\pi} \int_{-\infty}^\infty \exp{(i c K(\omega)t')} \exp{(-i\omega t)} \text{d} \omega \nonumber\\
    && = \frac{1}{2\pi} \sqrt{\frac{\pi}{\alpha_0 c t'}} \exp{\left( -\frac{(t-t')^2}{4 \alpha_0 c t'} \right)} \, \text{for} \phantom{X} t' > 0.
    \label{Eq:Attenuated_pressure_kernel_M}
 \end{eqnarray}   
\noindent
The factor $\omega/(cK(\omega))$ turns out to be approximately one for relevant frequencies and attenuation coefficients. This is shown for glycerine, a liquid with rather high viscosity, in the next subsection: even for a high frequency of 100 MHz, the factor is 0.994. For lower frequencies or lower acoustic attenuation than in glycerine this factor is even closer to one. Therefore, $M_p (t,t')$ is in a good approximation a Gaussian pulse, where the amplitude is reduced and the width is increased by a factor of $\sqrt{\alpha_0ct'}$. For a thin absorbing planar layer in an infinite medium, which can be modeled as a Dirac delta function in the thickness coordinate $z$, the virtual pressure $p_\text{virt}(z,t)$ for $z>0$ is proportional to $\delta(t-z/c)$. From Eq. (\ref{Eq:Attenuated_pressure_kernel_M}) one gets for the attenuated pressure wave
\begin{eqnarray}
    p(z,t)&& \varpropto \int_{-\infty}^\infty \delta(t'-z/c)M_p(t,t')dt' \thickapprox \nonumber \\
    && \thickapprox \frac{1}{2\pi} \sqrt{\frac{\pi}{\alpha_0 z}} \exp{\left( -\frac{(t-z/c)^2}{4 \alpha_0 z} \right)},
    \label{Eq:attenuated_pressure_wave}
\end{eqnarray} 
\noindent
which is the inverse Fourier transform (Eq. (\ref{subeq:2})) of $\exp{(iK(\omega)z)}$ \cite{Patch.2007}. This is also true for a general complex $K(\omega)$, e.g. with different power laws than the square of the frequency, such as for porcine fat tissue with a power law exponent of 1.5 (section \ref{sec:1Dfat}). More details on compensation of acoustic attenuation can be found in references \cite{Szabo.2014, LaRiviere.October23292005, LaRiviere.2006, Ammari.2012, DeanBen.2011, Kowar.2012, Burgholzer.2007b, Burgholzer.2010, Burgholzer.2010b, Burgholzer.2011, Treeby.2010, Treeby.2010b, Treeby.2013}. \par
Eq. (\ref{Eq:Attenuated_pressure_kernel_M}) can be discretized to produce a matrix equation
\begin{equation}
    \mathbf{p} = \mathbf{M}_p \mathbf{p}_\text{virt},
    \label{Eq:Matrix_pressure}
\end{equation}
\noindent
where $\mathbf{p}$ and $\mathbf{p}_\text{virt}$ are the vectors of the attenuated and virtual pressure signal at discrete time steps, respectively. $\mathbf{M}_p$ is the matrix at these time steps calculated from Eq. (\ref{Eq:Attenuated_pressure_kernel_M}).

\par
In a similar way, we have calculated a local relation between the actual temperature $T(\mathbf{r},t)$ and the virtual wave signal $T_\text{virt} (\mathbf{r},t)$ from Eq. (\ref{Eq:Virt_wave_equation}) and (\ref{Eq:Diffusion_equation}), or in frequency domain from the Helmholtz equations (\ref{Eq:Virtual_wave_frequency}) and (\ref{Eq:Temperature_frequency}) \cite{Burgholzer.2017}. Replacing $\omega$ by $c\sigma(\omega)$ in Eq. (\ref{Eq:Virtual_wave_frequency}) gives Eq. (\ref{Eq:Temperature_frequency}) by multiplying with $\frac{ic}{\alpha \sigma(\omega)}$ and identifying
\begin{equation}
    \tilde{T}(\mathbf{r},\omega) = \frac{ic}{\alpha \sigma(\omega)}\tilde{T}_\text{virt}(\mathbf{r},c\sigma(\omega)).
    \label{Eq:Temperature_virtual_wave_frequency}
\end{equation}
\noindent
The inverse Fourier transformation in Eq. (\ref{subeq:Inverse_fourier}) can be integrated analytically and can be written as \cite{Burgholzer.2017}
\begin{eqnarray}
    T(\mathbf{r},t) &&= \int_{- \infty}^\infty T_\text{virt}(\mathbf{r},t') M_T(t,t')\text{d}t', \nonumber \\
    && \text{with} \nonumber \\
    M_T(t,t') &&\equiv \frac{c}{\sqrt{\pi \alpha t}} \exp{\left( - \frac{c^2 t'^2}{4 \alpha t} \right)} \phantom{X} \text{for} \phantom{X} t>0.
    \label{Eq:Temperature_virtual_wave}
\end{eqnarray}
\noindent
Eq. (\ref{Eq:Temperature_virtual_wave}), like Eq. (\ref{Eq:Temperature_virtual_wave_frequency}) connects the temperature signal to the virtual wave signal at the same location $\mathbf{r}$, but in time domain instead of the temporal frequency domain.\par
For the example of a thin absorbing layer in an infinite medium the virtual temperature wave consists of two Dirac delta pulses running in the positive and negative $z$-direction proportional to $\frac{1}{2} \delta(z \pm ct) = \frac{1}{2c}\delta\left( t \pm \frac{z}{c} \right)$. From Eq. (\ref{Eq:Temperature_virtual_wave}) we get the 1D Green’s function
\begin{eqnarray}
    T(z,t)&& \varpropto \int_{-\infty}^\infty \frac{1}{2} \delta(z \pm ct') M_T(t,t')\text{d}t' \nonumber \\
    && = \frac{1}{\sqrt{4 \pi \alpha t}}\exp{\left(-\frac{z^2}{4 \alpha t}\right)}.
\end{eqnarray} 
\noindent
Eq. (\ref{Eq:Temperature_virtual_wave}) can be discretized to produce a matrix equation
\begin{equation}
    \mathbf{T} = \mathbf{M}_T \mathbf{T}_\text{virt},
    \label{Eq:Matrix_temperature}
\end{equation}
\noindent
similar to Eq. (\ref{Eq:Matrix_pressure})Eq. (\ref{Eq:Matrix_temperature}) and Eq. (\ref{Eq:Matrix_pressure}) can be inverted only with appropriate regularization, as the matrices $\mathbf{M}_T$ and $\mathbf{M}_p$ are rank deficient. The truncated-singular value decomposition (T-SVD) method is used in the following to get the reconstructed signal $\mathbf{T}_\text{rec}$ as an estimate for $\mathbf{T}_\text{virt}$ from the thermographic signal $\mathbf{T}$  or $\mathbf{p}_\text{rec}$ as an estimate for $\mathbf{p}_\text{virt}$ from the pressure signal $\mathbf{p}$, respectively. The truncation value for the smallest singular values is $1/\text{SNR}$, with SNR being the signal-to-noise ratio for the measured temperature or pressure.\par
Image reconstruction is now a two-stage process, where two successive inverse problems are solved. In photoacoustic imaging in a first step $p_\text{virt}$ from the measured pressure signal $p$ on the sample surface is determined and then in a second step the initial pressure distribution $p_0 (\mathbf{r})$ is determined by acoustic reconstruction methods, such as back-propagation or time reversal reconstruction algorithms \cite{Burgholzer.2007}. In photothermal imaging the initial temperature distribution $T_0 (\mathbf{r})$ is determined by the same reconstruction methods from the virtual temperature wave $T_\text{virt}$, which was calculated in the first step from the measured surface temperature $T$. This two-stage process for imaging can be used in 1D, 2D, and 3D. In one dimension the second step, the reconstruction by back-projection is rather trivial: the time signal of the virtual pressure or temperature wave multiplied by the constant sound speed $c$ gives directly the initial pressure or temperature at the depth $z=ct$. Therefore, in the beginning we will show a 1D example. 

\subsection{Compensation of acoustic attenuation for 1D photoacoustic imaging}
\label{sec:compensation_acoustic_attenuation}

In frequency domain, according to Eq. (\ref{Eq:pressure_frequency}) the amplitude of the wave component with frequency $\omega$ is damped by the factor $\exp{(-\alpha_0 \omega^2 z)}$ during propagation from depth $z$ to the sample surface. For  frequencies larger than the truncation frequency $\omega_\text{cut}$, the amplitude of these wave components is damped below the noise level, and at $\omega_\text{cut}$ we get
\begin{equation}
    \text{SNR} \exp{(-\alpha_0 \omega_\text{cut}^2 z)} = 1 \phantom{X} \text{or} \phantom{X} \omega_\text{cut} = \sqrt{\frac{\ln(\text{SNR})}{\alpha_0z}}.
    \label{Eq:Photoacoustic_cut_off}
\end{equation}
For the spatial resolution in photoacoustic imaging the minimal possible width of the acoustic signal in the time domain is essential. A small width enables high spatial resolution, which corresponds to a broad frequency bandwidth. If the frequency bandwidth is limited according to Eq. (\ref{Eq:Photoacoustic_cut_off}), the spatial resolution limit according to Nyquist is half the wavelength at this frequency
\begin{equation}
    \delta_\text{resolution} = \frac{\pi}{\omega_\text{cut}}c = \pi c \sqrt{\frac{\alpha_0 z}{\ln \text{(SNR)}}}.
    \label{Eq:delta_resolution}
\end{equation}

The resolution limit can be validated by the reconstruction of a thin absorbing planar layer in an infinite medium, which is modeled as a Dirac delta function in the thickness coordinate $z$, and where the virtual pressure $p_\text{virt} (z,t)$ for $z>0$ is proportional to $\delta(t-z/c)$. The Fourier transformation of a delta pulse is $\exp(i\omega z/c)$ and shows all frequencies with equal amplitude. The reconstruction in the attenuation limited frequency bandwidth is the inverse Fourier transformation, where the frequency integral is taken from $-\omega_\text{cut}$ to $+\omega_\text{cut}$, which gives
\begin{eqnarray}
    p_\text{rec}(z,t)&& = \frac{1}{2\pi} \int_{-\omega_\text{cut}}^{+\omega_\text{cut}} \exp\left( i \omega \frac{z}{c} \right) \exp(- i \omega t) \text{d}\omega \nonumber \\ \label{eq:recP}
    && = \frac{1}{\pi}\frac{1}{z/c-t} \sin\left( \omega_\text{cut} \left( \frac{z}{c}-t \right) \right) .
\end{eqnarray} 
This is a sinc-function, where the maximum at a distance $z$ is at $t=z/c$, which is the arrival time of the virtual wave. The zero points are at $t=\frac{z}{c} \pm \frac{\pi}{\omega_\text{cut}} =\frac{z\pm \delta_\text{resolution}}{c}$ with $\delta_\text{resolution}$ from Eq. (\ref{Eq:delta_resolution}). Compared to the measured pressure pulse without any compensation of attenuation given in Eq. (\ref{Eq:attenuated_pressure_wave}), the width of the reconstructed pressure pulse, Eq. (\ref{eq:recP}), is reduced by a factor of $\sqrt{\ln(\text{SNR})}$. A higher SNR allows a better resolution for the reconstruction.

\begin{table}
\caption{\label{tab:truncation_frequency} Truncation frequency and spatial resolution for glycerine.}
\begin{ruledtabular}
\begin{tabular}{ccc}
Propagation distance & Truncation frequency & Spatial resolution \\
$z$ & $\omega_\text{cut}/(2\pi)$ & $\delta_\text{resolution}$\\
\hline
1 mm & 42.9 MHz & 22.4 $\mu$m\\
5 mm & 19.2 MHz & 50.1 $\mu$m\\
10 mm & 13.6 MHz & 70.9 $\mu$m\\
\end{tabular}
\end{ruledtabular}
\end{table}

Glycerine is a liquid with rather high viscosity. Acoustic attenuation in glycerine is about 100 times higher than in water. At a temperature of 25$^\circ$ C and a frequency of 1 MHz the sound velocity is $c$ = 1923 m/s and the attenuation is 2.5 m$^{-1}$, which gives 22 dB/m or 0.22 dB/cm (multiplying the attenuation with 20 log$_{10}$($e$))  \cite{Shutilov.1988}. This gives $\alpha_0 = \alpha / \omega^2 = 2.5 / (2 \pi 10^6)^2$s$^2$/m$ = 6.3  \times 10^{-14} $s$^2$/m. Even for high acoustic attenuation as in glycerine and for high frequencies, e.g. up to 100 MHz, $\omega /(cK(\omega))$ can still be approximated by one in Eq. (\ref{Eq:attenuated_pressure_wave}), because $c\alpha_0 \omega = 0.076$ is small compared to one. \par
\begin{figure}
    \includegraphics[width=\columnwidth]{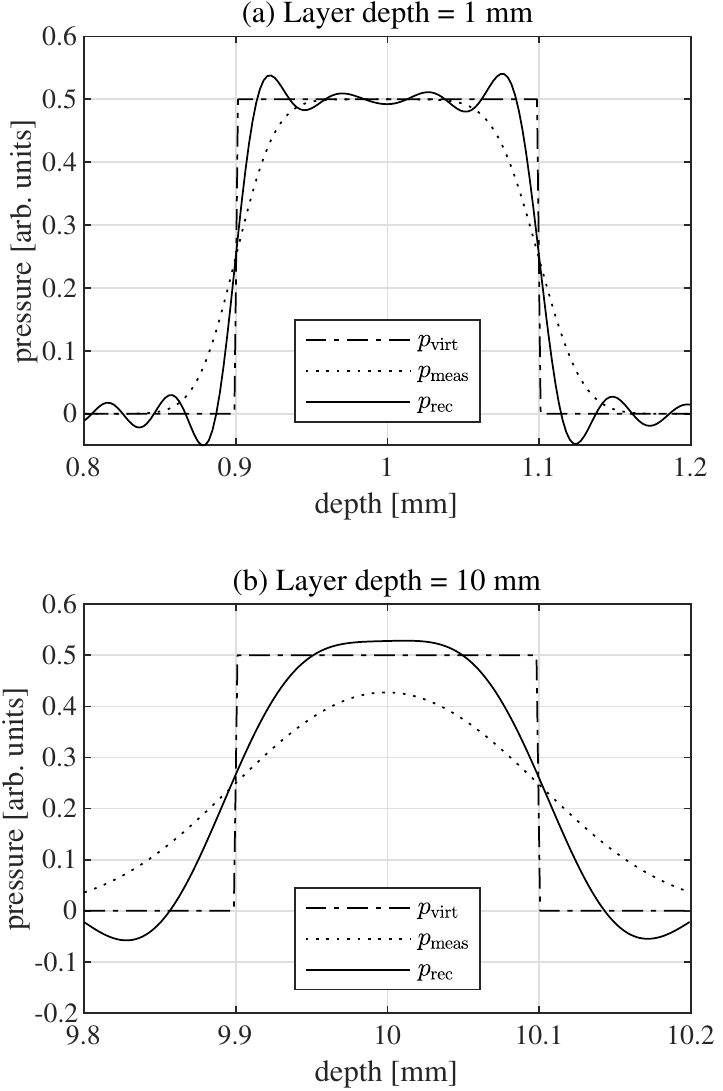}
    \caption{Two layers at a depth $z$ of 1 mm (a) and 10 mm (b), with a layer thickness of 0.2 mm. At a depth of 1 mm the attenuation of the measured pressure is significant lower. Therefore, also the reconstructed pressure $p_\text{rec}$ shows  a better spatial resolution. The theoretical resolution given in Table \ref{tab:truncation_frequency} is 22 $\mu$m around 1 mm depth and 71 $\mu$m around the depth of 10 mm.}
    \label{fig:Fig_pressure_depth_1mm}
\end{figure}

We will start with a simple 1D imaging problem: a thin planar layer at a certain depth $z$ below the surface of our detection plane. Due to acoustic attenuation the rectangular pressure pulse gets a different shape according to Eq. (\ref{Eq:Attenuated_pressure_kernel_M}), calculated by the discretized version of the matrix equation in (\ref{Eq:Matrix_pressure}). In our example, we take 3 layers at a depth $z$ of 1 mm, 5 mm, and 10 mm, with a layer thickness of 0.2 mm. Tab. \ref{tab:truncation_frequency} shows the truncation frequencies according to Eq. (\ref{Eq:Photoacoustic_cut_off}) and the resulting resolution limits from Eq. (\ref{Eq:delta_resolution}). Fig. \ref{fig:Fig_pressure_depth_1mm} shows the initial pressure distribution, which corresponds in 1D directly to the virtual pressure wave $\mathbf{p}_\text{virt}$, and the “measured” pressure signal $\mathbf{p}$, which was calculated from Eq. (\ref{Eq:virt_pressure_wave}), where 1~\% of Gaussian noise was added (SNR = 100). For image reconstruction, the inverse matrix was calculated by using the truncated singular value decomposition (T-SVD) \cite{Burgholzer.2017}. The SVD of the matrix $\mathbf{M}_p$ is a factorization of the form $\mathbf{U} \boldsymbol{\Sigma} \mathbf{V}^t$, where $\mathbf{U}$ and $\mathbf{V}$ are unitary matrices and $\boldsymbol{\Sigma}$ is a diagonal matrix with non-negative diagonal elements in decreasing order, called the singular values. For the pseudo-inverse matrix $\boldsymbol{\Sigma}^+$ the inverse of the non-vanishing diagonal elements in $\boldsymbol{\Sigma}$ is taken and the non-significant singular values were set to zero.. In the truncated SVD reconstruction, if the singular values get less than 1/SNR, they are set to zero. The pseudo-inverse of the matrix $\mathbf{M}_p$ is $\mathbf{M}_p^+ = \mathbf{V} \boldsymbol{\Sigma}^+ \mathbf{U}^t$. Then the reconstructed virtual pressure wave is
\begin{equation}
    \mathbf{p}_\text{rec} = \mathbf{M}_p^+\mathbf{p},
\end{equation}
\noindent
which is shown in Fig. \ref{fig:Fig_pressure_depth_1mm} around the depth of 1 mm and around the depth of 10 mm.

\subsection{Virtual wave concept for thermographic reconstruction}
\label{sec:VWC_1D}

In frequency domain, we can calculate the decay in amplitude of a thermal wave with frequency $\omega$ after a distance $z$ either by directly solving the Helmholtz equation (\ref{Eq:Temperature_frequency}) or by inserting the Ansatz \cite{Almond.1996} 
\begin{equation}
    T(z,t) = \Re \left(T_0 \exp \left(i \left(\sigma(\omega)z-\omega t\right)\right)\right) ,
\end{equation}
\noindent
with the complex wave number $\sigma(\omega) = \sqrt{\frac{i \omega}{\alpha}} \equiv \frac{1+i}{\mu} $ and a thin absorbing layer $T_0(z) = T_0 \delta(z)$ into the heat diffusion equation (\ref{Eq:Diffusion_equation}), which results in
\begin{equation}
    T(z,t) = \Re \left(T_0 \exp \left(-\frac{z}{\mu}\right)\exp\left( i \frac{z}{\mu} - i \omega t \right)\right), 
\end{equation}	
\noindent	
where $\Re$ is the real part and $\mu(\omega) \equiv \sqrt{2 \alpha / \omega} $ is defined as the thermal diffusion length \cite{Salazar.2006}. The amplitude of the thermal wave is reduced by a factor $1/e$ after propagation of that length. The wavenumber or spatial frequency is $k(\omega) \equiv 1 / \mu(\omega) = \sqrt{\omega / 2 \alpha}$. Similar as for acoustic attenuation described in Sec.~\ref{sec:compensation_acoustic_attenuation}, for  frequencies larger than the truncation frequency $\omega_\mathrm{cut}$ the amplitude of this wave components are damped below the noise level, with the truncation wavenumber $k_\mathrm{cut}$, and for the truncation diffusion length, frequency, or wavenumber we get
\begin{eqnarray}
    \text{SNR }&& \exp \left(-\frac{z}{\mu_\mathrm{cut}}\right) = 1, \, \text{or} \nonumber \\ \omega_\mathrm{cut} &&= 2 \alpha \left(\frac{\ln(\text{SNR})}{z}\right)^2, \, \text{or} \nonumber \\ k_\mathrm{cut} &&= \frac{\ln(SNR)}{z}.
    \label{Eq:kcut}
\end{eqnarray}

Similar to Eq. (\ref{Eq:delta_resolution}) the spatial resolution 
\begin{equation}
    \delta_\mathrm{resolution} = \frac{\pi}{k_\mathrm{cut}} = \frac{\pi z}{\ln(\text{SNR})}
    \label{Eq:delta_resolution_thermal}    
\end{equation}
\noindent
is half of the wavelength of the truncation frequency. For thermographic reconstruction, the resolution limit decays linear with depth z. In comparison, for photoacoustic reconstruction the resolution limit decays proportional to the square root of z (Eq. (\ref{Eq:delta_resolution})). \par
The objective of the solution of the inverse problem in thermographic reconstruction is to calculate the virtual wave temperature from the noisy temperature measurements. The one-dimensional solution of the wave equation is a wave package of constant shape travelling with sound velocity, because the group velocity is equal to the phase velocity. Therefore, in 1D the virtual wave immediately gives the reconstruction at time $t = 0$ by projecting it back at a distance $z = c t$. Similar to photoacoustic imaging, the optimum reconstruction can be achieved in the temporal frequency domain ($\omega$-space) by considering frequencies up to the truncation frequency $\omega_\mathrm{cut}$. If $T_\mathrm{virt}$ has the shape of a sharp pulse, ideally a Dirac delta distribution, the pulse will keep its shape over a travel distance, as the virtual wave equation (\ref{Eq:Virtual_wave_frequency}) has only solutions without dispersion and attenuation. In $\omega$-space, for $\tilde{T}_\mathrm{virt}$ also the high frequencies will contribute to the signal after some propagation. From the delta pulse in $T_\mathrm{virt}$, the related temperature signal $T$ gets broadened with increasing distance $z$ according to Eq. \ref{Eq:Temperature_virtual_wave}.\par
\begin{figure}
\includegraphics[width=\columnwidth]{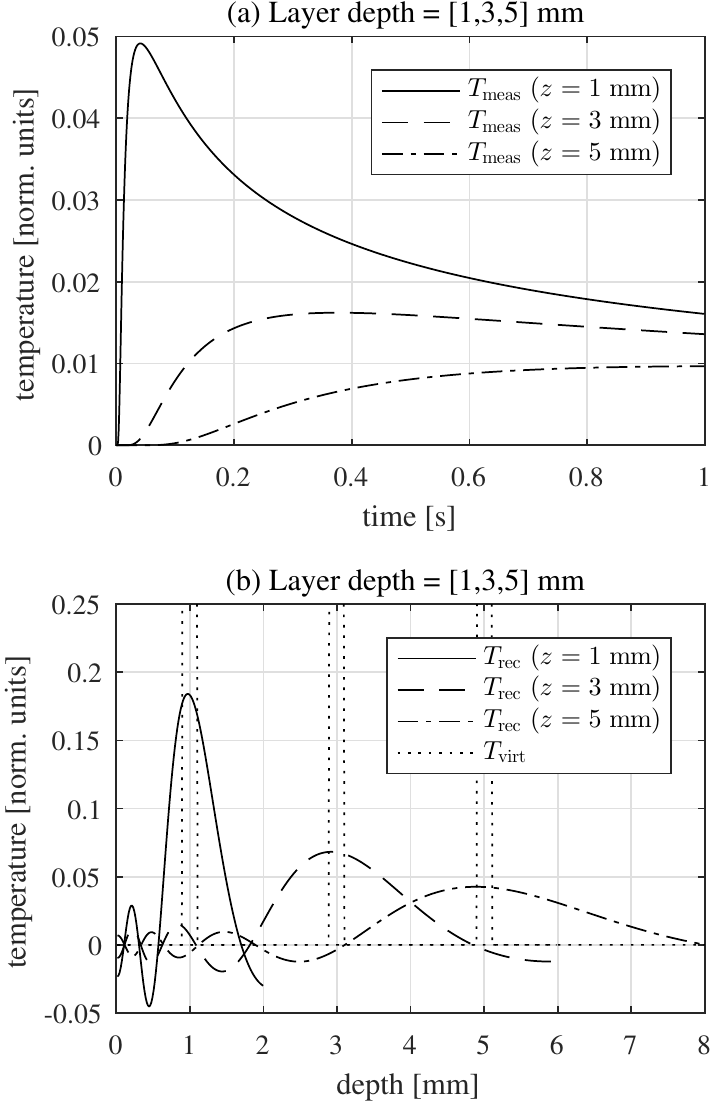}
\caption{\label{fig:Fig_temperature_depth} Thermal transients resulting from pulse heating of three different layer sources at a depth $z$ of 1 mm, 3 mm and 5 mm, with a layer thickness of 0.2 mm and the thermal properties of stainless steel. The reconstructed temperature signals from the virtual signals in three different depths and the decreasing resolution with travelling distance $z = c t'$ is shown in (b). The resolution is in the order of the layer depth.} 
\label{Fig:Temperature_Depth}
\end{figure}

In Fig. \ref{Fig:Temperature_Depth}(a) the time-dependent temperature signal calculated using Eq. (\ref{Eq:Matrix_temperature}) is shown for three different heat sources at layer depths of 1, 3 and 5 mm. The layer thickness is the same as in the photoacoustic example with 0.2 mm. A low carbon steel (<0.4 \%C) with the following material parameters is theoretically investigated: k = 43 W/(mK), C$_p$ = 465 J/(kgK) and $\rho$ = 7850 kg/m$^3$. The resulting thermal diffusivity, which describes the speed of heat propagation, is $\alpha = k / (\rho C_p)$ = 11.8 10$^{-6}$ m$^2$/s. The characteristic time (or time scale) to reach the maximum temperature in an 1D heat diffusive process is given by $t_\mathrm{D} = z^2 / (2\alpha)$. Due to this quadratic dependence, the temperature curves in Fig. \ref{Fig:Temperature_Depth}(a) show very widespread time scales for different layer depths $z$. This can be interpreted by the concept of "thermal waves": the short time behavior is dominated by the fastest propagating, high frequency components of the pulse and the long time behavior by the low frequency components, with a phase velocity of $\sqrt{2 \alpha \omega}$ \cite{Almond.1996}. The sample can be considered as a low pass filter, because the high frequency components are heavily attenuated as they propagate, and so only the slow moving low frequency components reach the surface from deeper layer sources in the sample. In Fig. \ref{Fig:Temperature_Depth}(b), the reconstructed signal $T_\mathrm{rec}$ is shown for the three rectangular signals of $T_\mathrm{virt}$, whereby for the regularization the truncated-SVD is used with a truncation value for the smallest singular value of 1 / SNR with a SNR of 100. The truncation frequencies are listed in Tab. \ref{tab:truncation_frequency2} for the three different layer depths. As the resolution limit is the size of a small, or ideally point-like, reconstructed source at a certain depth, the resolution after some propagation distance $z = c t'$, is the half the wavelength at this frequency or the width of the reconstructed signal calculated by the T-SVD. This is in good agreement with the spatial resolution in Eq. (\ref{Eq:delta_resolution_thermal}). The thermographic resolution limit is proportional to the travel distance $z$ of the signal between the excitation location and the detection surface and inversely proportional to the natural logarithm of the SNR, whereby the thermal diffusivity has no influence. The resolution limits for a SNR of 100 are given in Tab. \ref{tab:truncation_frequency2}. 

\begin{table}
\caption{\label{tab:truncation_frequency2} Truncation frequency and spatial resolution for steel in thermographic reconstruction with a SNR = 100.}
\begin{ruledtabular}
\begin{tabular}{ccc}
Propagation distance & Truncation frequency & Spatial resolution \\
$z$ & $\omega_\text{cut}/(2\pi)$ & $\delta_\text{resolution}$\\
\hline
1 mm & 79.5 Hz & 0.68 mm\\
3 mm & 8.8 Hz & 2.05 mm\\
5 mm & 3.2 Hz & 3.41 mm\\
\end{tabular}
\end{ruledtabular}
\end{table}

\section{Thermodynamics and Information}
\label{sec:thermodynamic}

The process of information gaining during photoacoustic and photothermal imaging is a physical one which has to obey the laws of (non–equilibrium) thermodynamics. Determining the virtual pressure or temperature wave from the measured pressure or temperature is an ill-posed or an ill-conditioned inverse problem, which needs regularization \cite{Hansen.1998}. In the previous section the  truncation frequency $\omega_\mathrm{cut}$ was chosen as regularization parameter, where it was assumed by regularization using the T-SVD method that all signal components below $\omega_\mathrm{cut}$ could be used for reconstruction, and all components above $\omega_\mathrm{cut}$, where the signal is damped below the noise level, provide no additional information for reconstruction. Therefore, the signal amplitude compared to noise level, plays an important role in determining the truncation frequency $\omega_\mathrm{cut}$, which serves as a regularization parameter for the inverse problem. Where does this noise come from and what is its relation to information gaining in imaging?\par
Before answering these essential questions some basics of thermodynamics and statistical physics should be reviewed. A microstate is a specific microscopic configuration of a thermodynamic system, e.g. with certain locations or velocities of all of the molecules. A microstate can be represented as a single point with coordinates $x$ in a usually very high dimensional phase space. In contrast, the macrostate of a system refers to a few macroscopic properties, such as its temperature, pressure, volume or density. A macrostate is characterized by a probability distribution $\rho(x)$ in the phase space of possible microstates. This distribution describes the probability of finding the system in a certain microstate. In classical mechanics, the position and momentum variables of a particle can vary continuously, so the set of microstates is actually uncountable and $x$ gets a continuous variable. The time evolution of all the particles is determined by the Hamiltonian $H$, which gives the total energy of a microstate, the sum of the kinetic and potential energy. The systems energy 
\begin{equation}
    U_t = \int H(x) \rho_t(x)\mathrm{d}x\,,
    \label{Eq:System_energy}
\end{equation}
at time $t$ is the mean value or also called expectation value of the system Hamiltonian $H$, where $\rho_t(x)$ is the probability distribution at time $t$.

\begin{figure*}
\includegraphics[width=0.8\textwidth]{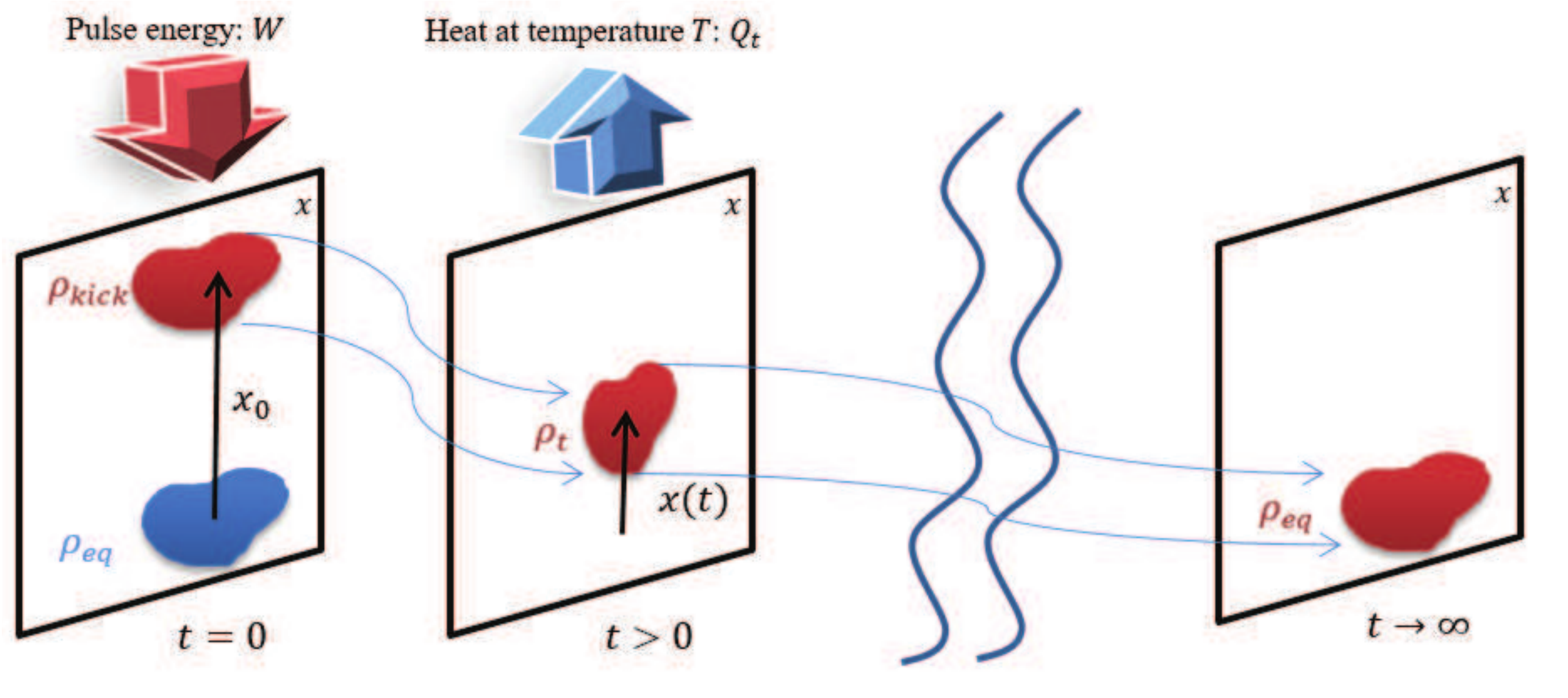}
\caption{Illustration of the thermodynamics of the imaging process in phase space: a system in equilibrium state $\rho_\mathrm{eq}$ is kicked at a time $t=0$ with magnitude $x_0$ and performed work $W$ to a state $\rho_\mathrm{kick}$ far from equilibrium, followed by a dissipative or diffusive process back to equilibrium. Till the time $t$ the heat $Q_t$ flows to the surroundings of the system at an ambient temperature $T$. At $t \rightarrow \infty$ equilibrium is reached again and $Q_t$ gets $W$. $x$ is the coordinate in phase space or a set of reduced variables which captures the information on the work (see text). The arrows connecting $\rho_\mathrm{kick}$ at time $t=0$, $\rho_t$ at $t>0$, and $\rho_\mathrm{eq}$ at $t \rightarrow \infty$ indicate the tube of trajectories, which is “thin” for macroscopic systems as deviations from the mean values $x(t)$ are small.}
\label{fig:Fig_energy_state}
\end{figure*}

Here, we follow the work of Esposito and van den Broeck \cite{Esposito.2011} about the second law and the connection between entropy production and information loss. They have used a general Hamiltonian system in contact with an ideal heat bath at temperature $T$. $W_t$ is the work performed on the system and $Q_t$ is the heat coming from the system into the ideal heat bath. In our case, the system is the sample, the performed work is the energy $W$ of the short laser pulse at the time $t=0$ and $Q_t$ is the diffused heat or the dissipated heat from the acoustic attenuation till time t. The ideal heat bath is the samples environment having an ambient temperature T (see Fig. \ref{fig:Fig_energy_state}).
The system entropy is the Shannon entropy
\begin{equation}
    S_t = -k_\mathrm{B} \int \rho_t(x) \ln(\rho_t(x)) \mathrm{d}x,
    \label{Eq:Shannon_entropy}
\end{equation}
\noindent
where $k_\mathrm{B}$ is the Boltzmann constant, and the corresponding system free energy is given by 
\begin{equation}
    F_t = U_t - T S_t.
    \label{Eq:Free_energy}
\end{equation}
\noindent
Here $T$ is the temperature in Kelvin of an ideal heat bath with which the system is in contact - the ambient temperature of the sample. The excitation laser pulse deposits the work $W$ at the time $t=0$, which either diffuses as heat out of the system or is the heat from dissipation of the photoacoustic pulse by acoustic attenuation, called $Q_t$ from the laser pulse till the time $t$. After a long time, the system is in the same thermal equilibrium as in the time before the excitation pulse, and all the energy $W$ of the laser pulse has left the system and is in the heat bath ($Q_t=W$), as sketched in Fig. \ref{fig:Fig_energy_state}. 

Following conservation of energy (first law of thermodynamics) the corresponding energy change $\Delta U_t := U_t-U_\mathrm{eq}$ of the system is given by
\begin{equation}
    \Delta U_t = W - Q_t \phantom{X} \text{for} \phantom{X} t>0.
    \label{Eq:Energy_change}
\end{equation}
\noindent
At $t \rightarrow \infty$ equilibrium is reached again, $Q_t$ gets $W$, and the change in system energy $\Delta U_t=0$.\par
If the system is in equilibrium with the surroundings at the ambient temperature $T$, the equilibrium distribution is the canonical distribution
\begin{eqnarray}
    \rho_\mathrm{eq}(x) &&= \frac{1}{Z}\exp \left(-\beta H(x)\right) \, \text{with} \nonumber \\ Z &&= \int \exp \left(-\beta H(x)\right) \text{d}x,
    \label{Eq:canonical_distribution}
\end{eqnarray}
\noindent
where the normalization function of the equilibrium distribution is called the canonical participation function $Z$ and the thermodynamic $\beta=1/(k_\mathrm{B} T)$. This Maxwell-Boltzmann distribution $\rho_\mathrm{eq}$ can be determined by minimizing the entropy $S_t$ under the two constraints on the total particle number and on the average energy per particle, e.g. \cite{Presse.2013, Gearhart.2001}.
For the equilibrium entropy one gets 
\begin{equation}
    S_\mathrm{eq} = k_\mathrm{B} \int \rho_\mathrm{eq}(x)\left( \ln Z + \beta H(x) \right) \text{d}x = k_\mathrm{B} \ln Z + \frac{U_\mathrm{eq}}{T},
\end{equation}
\noindent
using Eq. (\ref{Eq:Shannon_entropy}) and Eq. (\ref{Eq:System_energy}), and for the equilibrium free energy
\begin{equation}
    F_\mathrm{eq} = U_\mathrm{eq} - T S_\mathrm{eq} = -k_\mathrm{B}T \ln Z.
    \label{Eq:equilibrium_free_energy}
\end{equation}

According to the second law of thermodynamics the entropy of an adiabatically insulated system  increases monotonically until thermodynamic equilibrium is established, where the relative entropy $-k_\mathrm{B} D(\rho_t \mid \rho_\mathrm{eq})$ gets zero, with the Kullback-Leibler divergence
\begin{equation}
    D(\rho_t |\rho_\mathrm{eq}) = \int \rho_t(x) \ln \left( \frac{\rho_t(x)}{\rho_\mathrm{eq}(x)} \right) \text{d}x
    \label{Eq:Kullback_Leibler}
\end{equation}
\noindent
between the non-equilibrium distribution $\rho_t$ at time $t$ and the equilibrium distribution $\rho_\mathrm{eq}$ \cite{Cover.2006}. The Chernoff-Stein’s Lemma gives a precise meaning to $D(f\mid g)$ as a “distance” between two distributions: if $n$ data from $g$ are given, the probability of guessing incorrectly that the data come from $f$ is bound by the error $\epsilon = \exp(-n D(f \mid g))$, for $n$ large \cite{Cover.2006,Parrondo.2009}.
The free energy of a non-equilibrium state $\rho_t$ is higher than that of the equilibrium state by an amount equal to the temperature times the information  $I_t$ needed to specify the non-equilibrium state
\begin{eqnarray}
    T I_t \equiv k_\mathrm{B}T D (\rho_t | \rho_\mathrm{eq}) && = -T S_t + \int H(x) \rho_t(x) \text{d}x + k_\mathrm{B} T \ln Z \nonumber \\
    && = U_t - T S_t + k_\mathrm{B}T \ln Z \nonumber \\ &&= F_t - F_\mathrm{eq} \equiv \Delta F_t.
    \label{Eq:temperature_rec}
\end{eqnarray} 
\noindent
For the first equality, Eq. (\ref{Eq:Kullback_Leibler}), Eq. (\ref{Eq:Shannon_entropy}), and Eq. (\ref{Eq:canonical_distribution}) are used, and the second equation uses Eq. (\ref{Eq:System_energy}). The third equation uses the definition of the free energy in Eq. (\ref{Eq:Free_energy}) and for the equilibrium state in Eq. (\ref{Eq:equilibrium_free_energy}). Due to the excitation pulse at time $t=0$, the free energy jumps from $F_\mathrm{eq}$ to $F_\mathrm{kick}=F_\mathrm{eq}+W$, because the pulse shifts the distribution to a higher energy $\Delta U_\mathrm{kick}=W$ (Eq. \ref{Eq:Energy_change}), but does not change the entropy (Fig. \ref{fig:Fig_energy_state}), $S_\mathrm{kick}=S_\mathrm{eq}$.\par
The change in the non-equilibrium system entropy $\Delta S_t=S_t-S_\mathrm{eq}$ can be written as a reversible contribution to the heat flow $-Q_t/T$ and of an irreversible contribution, called entropy production 
\begin{equation}
    \Delta S_t = \Delta_i S_t - \frac{Q_t}{T}.
    \label{Eq:Delta_entropy}
\end{equation}
\noindent
Together with the definition of the free energy in Eq. (\ref{Eq:Free_energy}), the first law Eq. (\ref{Eq:Energy_change}), and Eq. (\ref{Eq:temperature_rec}) we obtain for the entropy production
\begin{eqnarray}
    T \Delta_i S_t &&= T \Delta S_t + Q_t = \Delta U_t - \Delta F_t + Q_t \nonumber \\
    &&= W - \Delta F_t = W - T I_t \phantom{X} \text{for} \phantom{X} t>0.
    \label{Eq:Entropy_production}
\end{eqnarray} 
\noindent
The free energy $\Delta F_t$ jumps at time $t=0$ from zero to $W$ and then decreases and gets zero, the information according to Eq. (\ref{Eq:temperature_rec}) makes the jump to $W/T$, and then decreases to zero. $\Delta I_t$ is defined as the information loss and according to Eq. (\ref{Eq:Entropy_production}) we get the very important result, that the entropy production
\begin{equation}
    \Delta_i S_t = \Delta I_t = W / T - I_t.
\end{equation}
\noindent
\textit{is equal to the information loss}.\par
For a macroscopic sample the fluctuations, which are the variance or the “noise” around the mean values $x(t)$ are small and the change of the “shape” of the distribution during time evolution shown in Fig. \ref{fig:Fig_energy_state} can be neglected. Therefore, the change of the Shannon system entropy can be neglected ($\Delta S_t \approx 0$ in Eq. (\ref{Eq:Delta_entropy})) and the change in free energy is equal to the change in system energy, $\Delta F_t \approx \Delta U_t$, which gives from Eq. (\ref{Eq:temperature_rec}) for the information
\begin{eqnarray}
    I_t &&= k_\mathrm{B}D(\rho_t \mid \rho_\mathrm{eq}) = \frac{\Delta F_t}{T} \approx \frac{\Delta U_t}{T} \nonumber\\
    &&= \frac{W-Q_t}{T} \approx k_\mathrm{B}D(\rho_\mathrm{kick} \mid \rho_\mathrm{eq}) - k_\mathrm{B}D(\rho_\mathrm{kick} \mid \rho_\mathrm{t}).
    \label{Eq:Change_free_energy}
\end{eqnarray} 
\noindent
In this very good approximation for macroscopic samples, the Kullback-Leibler divergence gets a distance in the mathematical sense, and the energy $W-Q_t$, which has not yet dissipated or diffused to the surroundings at temperature $T$, divided by this temperature is the information content. After a long time, when all the work $W$ has dissipated or diffused, $I_t$ gets zero and the equilibrium distribution $\rho_\mathrm{eq}$ is reached. But according to Chernoff-Stein’s Lemma after some time $t_\mathrm{cut}$ the distribution $\rho_{t_\mathrm{cut}}$ cannot be statistically distinguished from $\rho_\mathrm{eq}$, if $k_\mathrm{B} T D(\rho_t \mid \rho_\mathrm{eq} )$ gets smaller than the equilibrium energy $U_\mathrm{eq}$.\par
For real world examples having a high dimensional phase space $x$ the actual distribution density $\rho_t$ can be hardly determined, but for the information loss from Eq. (\ref{Eq:Change_free_energy}) it is sufficient to evaluate the dissipated or diffused work from the mean value equations. Usually, a set of reduced variables with a drastically lower dimensionality than the phase space can be used, which captures the information on the mean dissipated or diffused work. In the macroscopic approximation the actual noise or fluctuation around the mean value is irrelevant. Only its amplitude is significant for setting the lower bound in Chernoff-Stein’s Lemma. The following subsections give a small example for dissipation and for diffusion to clarify the relation between entropy production and information loss. 

\subsection{Kicked Brownian Particle}
\label{sec:Brownian_particle}

We now consider the velocity $v$ of a particle as a stochastic process, which was used to describe Brownian motion of a particle, but for simplicity only one velocity component (1D) is considered here. The essential relation between fluctuation (noise) and dissipation (viscous damping) is shown. Stochastic processes can be described mathematically, e.g. by Master equations, Langevin- or Fokker-Planck equations, shown e.g. in the books of van Kampen \cite{vanKampen.2007}, Gardiner \cite{Gardiner.1985}, or Risken \cite{Risken.19891996printing}. In the Langevin-equation 
\begin{equation}
    \frac{\text{d}v(t)}{\text{d}t}=-\gamma v(t) + \sigma \eta (t).
    \label{Eq:Langevin}
\end{equation}
\noindent
the environmental forces on a particle in Newton’s law are a linear damping term together with random noise. The linear damping $-\gamma v$ is a viscous drag and $\sigma$ is the amplitude of the white noise $\eta$, which has a zero mean value and is uncorrelated in time: $\langle \eta (t)\eta (t') \rangle = \delta (t-t')$. The Langevin equation governs an Ornstein-Uhlenbeck (O-U) process, named after L. S. Ornstein and G. E. Uhlenbeck, who formalized the properties of this continuous Markov process \cite{Uhlenbeck.1930}. It was shown by using the statistical properties and the continuum limit of the white noise $\eta$, that the Langevin equation is equivalent to a description based on a Fokker-Planck equation. For the time-dependent distribution density $\rho_t (v)$ of the velocity a linear differential equation 
\begin{equation}
    \frac{\partial \rho_t(v)}{\partial t} = \frac{\partial(\gamma v \rho_t(v))}{\partial v} + \frac{\sigma^2}{2} \frac{\partial^2 \rho_t (v)}{\partial v^2}.
    \label{Eq:Langevin_Fokker_Planck}
\end{equation}
\noindent
can be derived\cite{Klages.2013, Risken.19891996printing}.\par
We start with the equilibrium state (zero time derivative if inserted into Eq. (\ref{Eq:Langevin_Fokker_Planck})), as the initial velocity distribution 
\begin{equation}
    \rho_\mathrm{eq}(v) = \frac{1}{Z} \exp\left(-\frac{\gamma}{\sigma^2}v^2\right) \phantom{X} \text{with} \phantom{X} Z = \sigma \sqrt{\frac{\pi}{\gamma}},
\end{equation}
\noindent
which is a Gaussian distribution with mean value zero and a variance $\text{Var}(v)=\sigma^2/(2\gamma)$, representing the statistical “noise”. Comparison with Eq. (\ref{Eq:canonical_distribution}) and the kinetic energy $H(v)=mv^2/2$ gives
\begin{equation}
    \beta H(v) = \frac{1}{k_\mathrm{B}T}\frac{mv^2}{2} = \frac{\gamma}{\sigma^2}v^2 \phantom{X} \text{or} \phantom{X} \text{Var}(v) = \frac{\sigma^2}{2 \gamma} = \frac{k_\mathrm{B}T}{m}.
    \label{Eq:FDtheorem}    
\end{equation}
\noindent
This relation states a connection between the strength of the fluctuations, given by $\sigma$, and the strength of the dissipation $\gamma$. This is the \textit{fluctuation-dissipation theorem} for uncorrelated white noise.\par
At a time zero the particle is kicked, which causes an immediate change in velocity of $v_0$ (kick magnitude) and the distribution density after the kick is
\begin{equation}
    \rho_\mathrm{kick}(v) = \rho_\mathrm{eq}(v-v_0)=\frac{1}{Z}\exp\left(-\frac{\gamma}{\sigma^2}(v-v_0)^2\right).
\end{equation}

\begin{figure}
\includegraphics[width=\columnwidth]{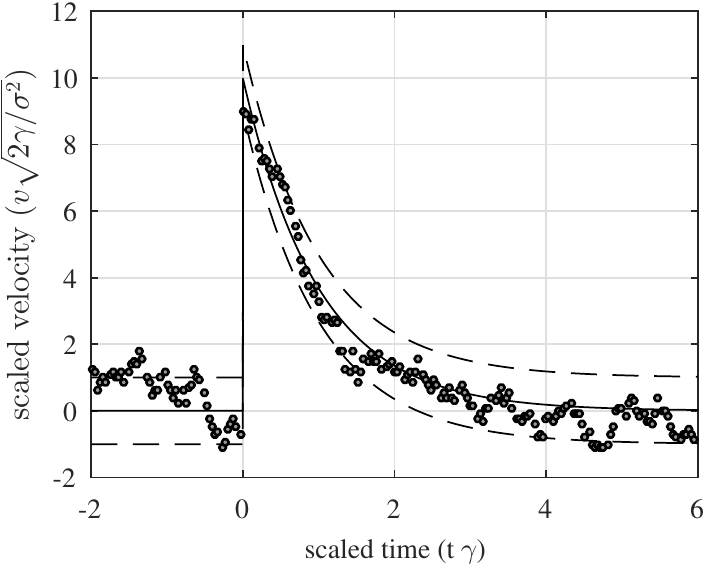}
\caption{The circles show a typical realization of a kicked Ornstein-Uhlenbeck process defined by the Langevin equation (\ref{Eq:Langevin}).The scaled time $t\gamma$ is on the horizontal axis. The velocity $v$ on the vertical axis is scaled to have a unit variance. At the time $t = 0$ a kick magnitude of $v_0=10$ is added to the scaled velocity. Increasing in time the information about the magnitude of the kick gets more and more lost due to the fluctuations. The solid line represents the mean, and the dashed lines the mean $\pm$ standard deviation, which is the square root of the variance.} 
\label{Fig:Fig_Kicked_particle}
\end{figure}

The solution of Eq. (\ref{Eq:Langevin_Fokker_Planck}) for the time-dependent distribution density $\rho_t (v)$ at $t > 0$ with $\rho_\mathrm{kick} (v)$ as initial condition at $t = 0$ is
\begin{eqnarray}
    \rho_t(v) &&= \frac{1}{Z}\exp \left(-\frac{\gamma}{\sigma^2}(v-\bar{v}(t))^2\right) \phantom{X} \text{with}  \nonumber \\  \bar{v}(t) &&= v_0 \exp(-\gamma t) \, \text{ for } \, t>0. 
    \label{Eq:distribution_density}
\end{eqnarray} 
\noindent
which gives a Gaussian distribution with time dependent mean value $\bar{v}(t)$ but a constant variance $\text{Var}(v)=\sigma^2/(2\gamma)$. As shown by Burgholzer and Hendorfer, this is a general feature of Gauss-Markov processes, also for higher dimensions: taking the equilibrium as an initial condition results for all times after the kick in a distribution with a constant (co)variance (matrix) equal to the equilibrium variance \cite{Burgholzer.2013}. One realization of the kicked O-U process is shown in Fig. \ref{Fig:Fig_Kicked_particle}. At the time $t = 0$ a kick with a magnitude of $v_0=10$ occurs. For a time $t > 0$ the information about the magnitude of the kick gets more and more lost due to the fluctuations. Now, this increasing information loss is quantified and compared to the mean entropy production.\par
According to Eq. (\ref{Eq:System_energy}) and (\ref{Eq:Energy_change}) due to the kick at time zero the energy of the particle is increased by $W = m v_0^2 / 2$ and
\begin{eqnarray}
    \Delta U_t &&= \int H(v)\rho_t(v)\text{d}v - \int H(v) \rho_\mathrm{eq}(v) \text{d}v \nonumber\\
    &&= \frac{m\bar{v}(t)^2}{2} = W \exp{(-2\gamma t)}
\end{eqnarray} 
\noindent
From Eq. (\ref{Eq:Shannon_entropy}) the entropy 
\begin{equation}
    S_t = - k_\text{B} \int \rho_t(v) \ln (\rho_t(v))\text{d}v = k_\mathrm{B} \ln Z + \frac{k_\mathrm{B}}{2}
\end{equation}
\noindent
stays constant in time ($\Delta S_t=0$), as the distribution is shifted in time but with a constant “shape”. Therefore the approximation in Eq. (\ref{Eq:Change_free_energy}) gets exact and the information content is
\begin{eqnarray}
    I_t &&= k_\mathrm{B} D (\rho_t \vert \rho_\mathrm{eq}) = k_\mathrm{B} \int \rho_t(v) \ln  (\rho_t(v)/\rho_\mathrm{eq}(v)) \text{d}v \nonumber \\
    &&= k_\mathrm{B} \frac{\gamma}{\sigma^2}\bar{v}(t)^2 = \frac{k_\mathrm{B}}{2}\frac{v_0^2}{\text{Var}(v)}\exp(-2\gamma t).
    \label{Eq:Information_exact}
\end{eqnarray}

According to Chernoff-Stein’s Lemma, if the information $I_t$ at a time $t_\mathrm{cut}$ gets less than a certain limit, here chosen as the zero-point energy divided by the temperature ($U_\mathrm{eq}/T=k_\mathrm{B}/2$), the distribution $\rho_t (v)$ cannot be statistically distinguished from the equilibrium distribution $\rho_\mathrm{eq} (v)$. The information content is
\begin{eqnarray}
    I_{t_\mathrm{cut}} = k_\mathrm{B}D(\rho_{t_\mathrm{cut}}\vert \rho_\mathrm{eq}) &&= \frac{k_\mathrm{B}}{2}\frac{v_0^2}{\text{Var}(v)}\exp{(-2 \gamma t_\mathrm{cut})} \nonumber \\
    &&= \frac{U_\mathrm{eq}}{T} = \frac{k_\mathrm{B}}{2},
    \label{Eq:Information}
\end{eqnarray}
\noindent
which gives
\begin{equation}
    t_\mathrm{cut} \gamma = \ln \left(\frac{v_0}{\sqrt{\text{Var}(v)}} \right) \equiv \ln (\text{SNR}).
\end{equation}
\noindent
At this truncation time $t_\mathrm{cut}$, not only the distribution cannot be distinguished from equilibrium, but also the mean velocity $\bar{v}(t)=v_0 \exp(-\gamma t)$ gets less than the noise level $v_0/\text{SNR}$. In Fig. \ref{Fig:Fig_Kicked_particle}, this is at the scaled time $t_\mathrm{cut}\gamma = \ln(10) \approx 2.3$. This coincidence, that the information related criterion in Eq. (\ref{Eq:Information}) gives the same time $t_\mathrm{cut}$ when the signal gets less than the noise always happens, if \textit{the energy described by the Hamiltonian $H$ is proportional to the square of the signal amplitude.} 

\subsection{Ideal gas diffusion between two boxes}
\label{sec:ideal_gas}

\begin{figure}
\includegraphics[width=0.8\columnwidth]{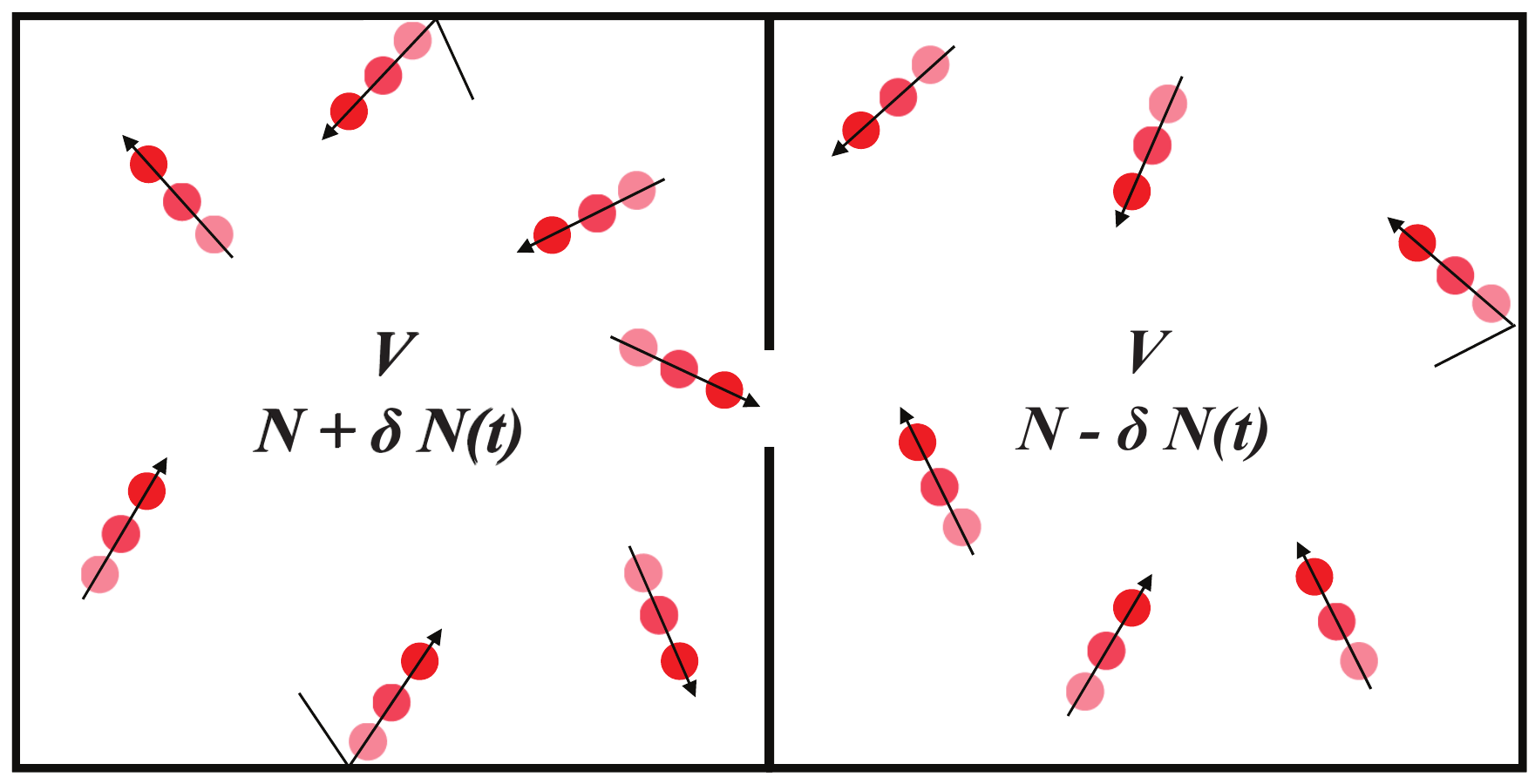}
\caption{Distribution of $2N$ particles between two boxes of volume $V$, which are connected by a small hole where particles can change place between the two boxes at a constant time rate $\gamma$ (effusion). At a certain time $t$, $N + \delta N(t)$ particles are situated in the left box and $N - \delta N(t)$ particles are in the right box.} 
\label{Fig:Fig_gas_boxes}
\end{figure}

Another example for an Ornstein-Uhlenbeck process using diffusion instead of dissipation is the distribution of $2N$ particles of an ideal gas between two boxes of volume $V$, which are connected by a small hole where particles can slip through between the two boxes (effusion). The constant time rate $\gamma$ is the reciprocal value of the mean time of a particle to stay in one of the boxes before it slips through the hole. At a certain time $t$, $N + \delta N(t)$ particles are situated in the left box and $N - \delta N(t)$ particles are in the right box (Fig. \ref{Fig:Fig_gas_boxes}).

$\delta N(t)$ can be approximated by an Ornstein-Uhlenbeck process for a big number of particles (see e.g. the tutorial introduction to stochastic processes by Lemons \cite{Lemons.2002}). At the time $t=0$, $N_0$ particles are “pumped” from the right to the left box. The mean value $\delta \bar{N}(t) = N_0 \exp(-\gamma t)$  for $t>0$, is given by an exponential decay in time, and the variance of the Gauss distribution is again constant in time with $\text{Var}(\delta N) = \frac{\sigma^2}{2 \gamma}$, according to Eq. (\ref{Eq:distribution_density}). However, $\text{Var}(\delta N)$ is not an energy, and therefore $\sigma^2$ cannot be expressed in terms of a rate $\gamma$ and a temperature T by requiring the equipartition of energy at equilibrium. The fluctuation - dissipation theorem as in Eq. (\ref{Eq:FDtheorem}) does not apply; $\text{Var}(\delta N)$ fluctuates without dissipation\cite{Lemons.2002}. The information loss can be described analogously to Eq. (\ref{Eq:Information_exact}) by an exponential decay 
\begin{equation}
    I_t = k_\mathrm{B} \frac{\gamma}{\sigma^2}\delta \bar{N}(t)^2 = \frac{k_\mathrm{B}}{2}\frac{N_0^2}{\text{Var}(\delta N)} \exp (-2\gamma t).
    \label{Eq:Information_loss_exponential}
\end{equation}

Spatial diffusion is the cause for the entropy production and is described according to Boltzmann by $S = k_\mathrm{B} \ln W$. $W$ is the number of possibilities to distribute the particles into the two boxes for a certain $\delta \bar{N}$. Using Stirlings formula for a big number of particles $N$ one gets for the entropy \cite{Presse.2013}
\begin{equation}
    \Delta S_t = -k_\mathrm{B} \frac{1}{N}\delta \bar{N}(t)^2 = -I_t.
\end{equation}
\noindent
By comparison with Eq. (\ref{Eq:Information_loss_exponential}) one gets again a relation between the fluctuations, given by $\sigma$, and the strength of the diffusion, given by the rate $\gamma$ (fluctuation-“diffusion” theorem) and for the variance of $\delta N$ we get
\begin{equation}
    \text{Var}(\delta N) = \frac{\sigma^2}{2 \gamma} = \frac{N}{2}. 
\end{equation}

\section{Imaging as an inverse problem}
\label{sec:imaging}

In the previous section, imaging with an excitation pulse at time $t=0$ has been described by means of nonequilibrium statistical physics. Pressure or temperature are proportional to the momentum or the kinetic energy of many particles, which fluctuate around their mean value. For macroscopic samples these fluctuations are very small and can usually be neglected in the thermodynamic limit. But for inverse problems these fluctuations are highly “amplified”, which was shown in the previous section for a system kicked out of the equilibrium, followed by a dissipative process back to equilibrium (Fig. \ref{fig:Fig_energy_state}). The inverse problem of estimating the kick-magnitude from an intermediate state a certain time after the kick is ill-posed. Just after the kick its magnitude can be estimated very well. A long time after the kick the state is nearly in equilibrium and all the information about the kick magnitude is lost. For macroscopic systems it could be shown that the information loss is in a good approximation just the mean dissipated energy or diffused heat divided by the temperature, which is the mean entropy production. In imaging, the spatial resolution and the information content are strongly correlated, and therefore a loss of information results in a loss of resolution, which is quantified for 1D examples in sections \ref{sec:compensation_acoustic_attenuation} and \ref{sec:VWC_1D}.\par
In sections \ref{sec:Brownian_particle} and \ref{sec:ideal_gas}, the mean dissipated energy and the resulting loss of information is calculated explicitly for a kicked Brownian particle and for an ideal gas diffusing between two boxes. For these study cases also the time-dependent probability distribution could be calculated explicitly, and the deduced information losses are in agreement with the results from mean value calculations, even when the distributions are broad and still far away from the thermodynamic limit. A truncation time could be given, for which the state after that time cannot be distinguished from the equilibrium distribution according to the Chernoff-Stein’s Lemma, when the Kullback-Leibler divergence gets too small. The amplitude of the fluctuations and the dissipation or the diffusion are not independent. Fluctuations and mean entropy production can be thought as two sides of the same coin and are connected by the fluctuation – dissipation theorem. For systems near thermal equilibrium in the linear regime such relations between entropy production and fluctuation properties have been found by Callen \cite{Callen.1952}, Welton \cite{Callen.1951}, and Greene \cite{Greene.1952}. This fluctuation-dissipation theorem is a generalization of the famous Johnson \cite{Johnson.1928} - Nyquist \cite{Nyquist.1928} formula in the theory of electric noise. It is based on the fact that in the linear regime the fluctuations decay according the same law as a deviation from equilibrium following an external perturbation.\par
For macroscopic systems it is not necessary to describe the full stochastic process to get the influence of the fluctuations. The relevant information loss can be calculated from the averaged behavior (mean value), which describes the usually known macroscopic evolution of the system in time, as the mean dissipated work or diffused heat divided by the temperature. This remarkable feature might be the reason that regularization methods for ill-posed inverse problems work so well, although they use only the mean value equations and not the detailed stochastic process to describe the time evolution. The choice of an adequate regularization parameter, e.g. the truncation value for the truncated singular value decomposition (SVD) method, is equivalent to the choice of the error level $\epsilon$ in the Chernoff-Stein’s Lemma.\par
A prominent class of ill-posed inverse problems is non-destructive imaging (NDI), where the information about the spatial pattern of a sample’s interior has to be transferred to the sample surface by certain waves, e.g., ultrasound or thermal waves (Fig. \ref{fig:Fig_Imaging_structures}). Imaging is done by reconstruction of the interior structure from the signals measured on the sample surface, e.g., by back-projection or time-reversal for a photo-acoustically induced ultrasound wave \cite{Burgholzer.2007, Burgholzer.2009}. There are several effects which limit the spatial resolution for photoacoustic imaging. Beside insufficient instrumentation and data processing one principle limitation comes from attenuation of the acoustic wave or from heat diffusion (section \ref{sec:Photothermal_acoustic_model}). Attenuation during wave propagation can be caused by dissipation of acoustic energy to heat or by acoustic scattering. In this article it is assumed that, as with dissipation, the information is also lost with scattering and that scattered waves are not used for image reconstruction. Using scattered waves for reconstruction would be possible, but this is beyond the scope of this tutorial. Therefore, the entire information loss due to acoustic attenuation is lost and cannot be compensated by subsequent processing algorithms. In this section it is shown that the loss of information, which is equal to the entropy production (section \ref{sec:thermodynamic}) as the dissipated energy or diffused heat divided by the temperature, is a principle thermodynamic limit, which cannot be compensated. Using the information loss and entropy production for a kicked process derived in section \ref{sec:thermodynamic} it is shown that the resolution limit depends just on the macroscopic mean-value equations and is independent of the actual stochastic process, as long as the macroscopic equations describe the mean work and therefore also the mean dissipated work.\par
In this section we do not attempt to model the process of acoustic attenuation or heat diffusion as a stochastic process, which we have done earlier as a Gaussian process\cite{Burgholzer.2013}. We use from section~\ref{sec:Photothermal_acoustic_model} the mean-value equations for the pressure $p(\mathbf{r},t)$ or temperature $T(\mathbf{r},t)$ evaluation in the frequency domain, $\tilde{p}(\mathbf{r},\omega)$ or $\tilde{T}(\mathbf{r},\omega)$, respectively.\par
For an acoustic wave in frequency domain, according to Eq. (\ref{Eq:pressure_frequency}) the amplitude of the wave component with frequency $\omega$ is damped by the factor $\exp(-\alpha_0 \omega^2 r)$ after propagating a distance $r$. The energy $\Delta U_\omega$ of the acoustic wave with frequency $\omega$ is proportional to the square of the pressure amplitude
\begin{equation}
    \Delta U_\omega = 0.5 \chi \Delta V \left|\tilde{p}(r,\omega)\right|^2 = 0.5 \chi \Delta V \exp (-2\alpha_0 \omega^2 r),
    \label{Eq:energy_pressure}
\end{equation}
\noindent
which can be found e.g. in Morse and Ingard \cite{Morse.19861968}, where $\chi = 1/(c^2 \rho)$ is the adiabatic compressibility with the density $\rho$ and $\Delta V$ is the measurement volume. Using that $\text{Var}(\rho) = k_\mathrm{B} T / (\chi \Delta V)$ is the variance of the pressure (e.g. from Landau and Lifshitz \cite{Landau.1980}) one gets from Eq. (\ref{Eq:Change_free_energy}) for the information content of the Fourier component with frequency $\omega$: $I_\omega = \Delta U_\omega / T = 0.5 k_\mathrm{B} \text{SNR}^2 \exp(-2 \alpha_0 \omega^2 r)$. The signal-to-noise ratio SNR at the distance $r=0$ is the reciprocal value of the square root of the variance of the pressure, as the signal amplitude in frequency domain is normalized to one (Eq. (\ref{Eq:pressure_frequency})). The truncation frequency $\omega_\mathrm{cut}$ is defined in the manner that the information content for frequencies larger than $\omega_\mathrm{cut}$ is so low that its distribution cannot be distinguished from the equilibrium distribution within a certain statistical error level (Chernoff-Stein’s Lemma). This level is $\Delta U_\mathrm{eq} / T = 0.5 \chi \Delta V \text{Var}(p)/T = 0.5 \chi \Delta V k_\mathrm{B} T / (\chi \Delta V)/T = 0.5 k_\mathrm{B}$, which results in
\begin{equation}
    I_{\omega_\mathrm{cut}} = 0.5 k_\mathrm{B} \text{SNR}^2 \exp (-2 \alpha_0 \omega_\mathrm{cut}^2 r) = 0.5 k_\mathrm{B}.
    \label{Eq:Information_reltated_criterion}
\end{equation}
\noindent
This gives the same truncation frequency as in Eq. (\ref{Eq:Photoacoustic_cut_off}), where the amplitude is damped just below the noise level. As described already in section \ref{sec:Brownian_particle}, the information related criterion in Eq. (\ref{Eq:Information_reltated_criterion}) gives the same truncation value as when the signal amplitude gets less than the noise, if the energy $\Delta U_\omega$ is proportional to the square of the signal amplitude $\left|\tilde{p}(r,\omega)\right|$ (Eq. (\ref{Eq:energy_pressure})).\par 
Similarly, also for particle diffusion the information content $I_t$ is proportional to the square of the mean difference in the particle number $\delta \bar{N}(t)$ (Eq. (\ref{Eq:Information_loss_exponential}). For heat diffusion and thermal waves the information content is proportional to the square of the temperature deviation from the equilibrium temperature, in real time space and in the frequency domain. Heat diffusion can be described as an Ornstein-Uhlenbeck process\cite{Burgholzer.2013,Groot.1984}. Therefore, the information related criterion in Eq. (\ref{Eq:Change_free_energy}) for thermal waves gives the same truncation frequency $\omega_\mathrm{cut}$ when the signal gets less than the noise with the results given in section \ref{sec:VWC_1D}.\par
As mentioned at the end of section \ref{sec:Imaging_model}, imaging in two (2D) or three dimensions (3D) can always be reduced using a two-stage process: first, for each detector location the virtual pressure signal in the absence of attenuation or the virtual thermal wave is calculated from the measured acoustic or thermal signal, respectively. This is a one-dimensional (1D) reconstruction problem, described in discrete time steps by Eq. (\ref{Eq:Matrix_pressure}) and Eq. (\ref{Eq:Matrix_temperature}). In a second step, any reconstruction method for photoacoustic tomography without acoustic attenuation, such as time-reversal or backprojection, can be used for reconstructions in higher dimensions \cite{Burgholzer.2007}. In 1D the virtual wave immediately gives the reconstruction at time $t=0$ by projecting it back at a distance  $ct$ (see section \ref{sec:compensation_acoustic_attenuation} and \ref{sec:VWC_1D}). Therefore, it is sufficient to examine the acoustic attenuation or thermal diffusion of 1D waves and the reconstruction in 1D. Compensation of acoustic attenuation or thermal diffusion in higher dimensions can always be reduced to 1D, which was also explicitly shown for signals from a layer (1D), cylinder (2D), and a sphere (3D) \cite{Burgholzer.2010b}. Therefore, in the beginning we will show a 1D example.

\subsection{1D imaging}
\label{sec:1Dimaging}

\subsubsection{Acoustic attenuation in a fat tissue layer}
\label{sec:1Dfat}

\begin{figure}
\includegraphics[width=\columnwidth]{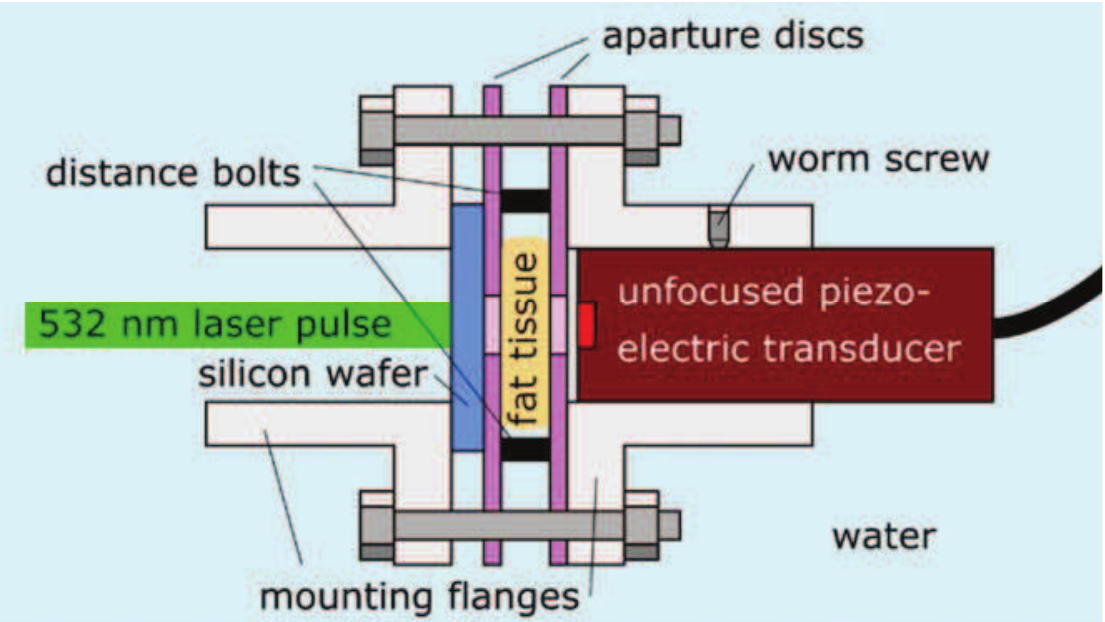}
\caption{Setup for the generation and detection of acoustic plane waves. Abrupt local heating of a silicon wafer by nanosecond laser pulses leads to the emission of strong broadband ultrasonic plane waves. Porcine subcutaneous fat tissue in the propagation path induces frequency dependent attenuation of the acoustic signals. The fat tissue is fastened between two aperture disks applying a small axial force on the tissue. Distance bolts with 6 mm or 20 mm ensure two precise lengths of the attenuation path. For these two lengths, attenuated acoustic plane waves were detected by an unfocused piezoelectric transducer, which was aligned by worm screws to ensure a one-dimensional signal propagation and detection. Image from Burgholzer et al.\cite{Burgholzer.2019} was edited and is used under CC BY 4.0.} 
\label{Fig:Fig_Setup_plane_waves}
\end{figure}

\begin{figure}
\includegraphics[width=\columnwidth]{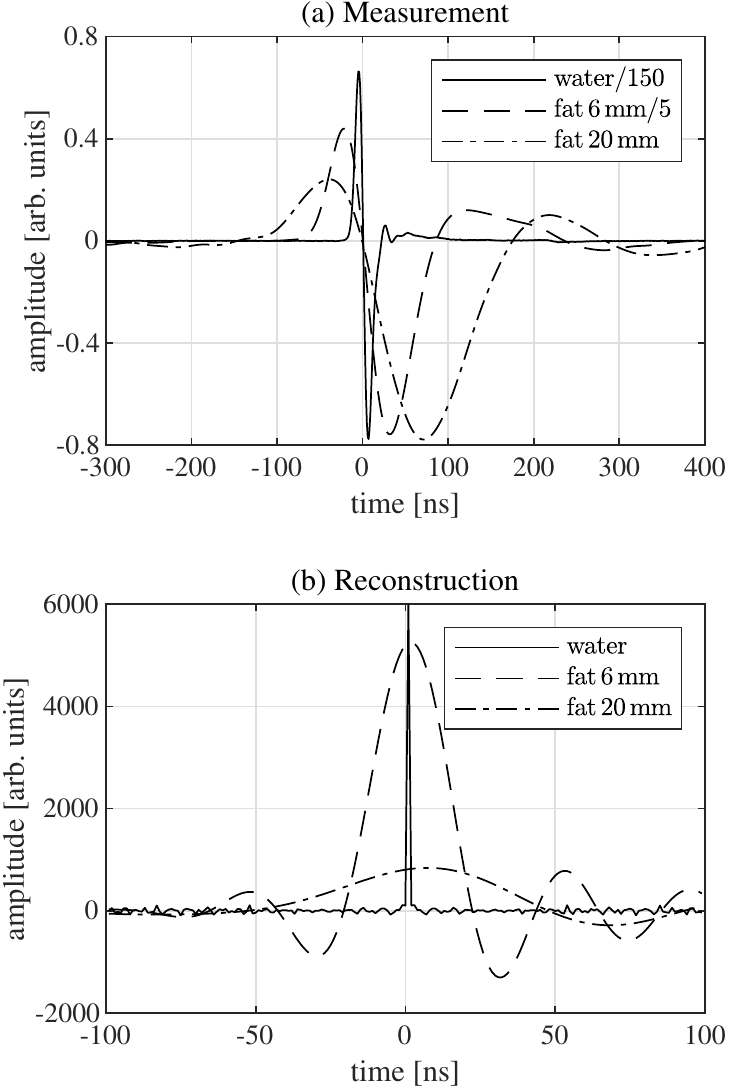}
\caption{Acoustic attenuation in a fat tissue layer: (a) Measured acoustic pressure as a function of time without fat tissue (“no fat”), with 6 mm thick porcine fat tissue and with 20 mm porcine fat tissue. To make comparison easier, the signals were time-shifted and scaled and (b) the reconstruction results using T-SVD for regularization to compensate attenuation in fatty tissue of 6 mm thickness and 20 mm thickness. This corresponds to a spatial resolution limit of 32 $\mu$m for 6 mm fat and 70 $\mu$m for 20 mm fat, resulting from entropy production. The matrix $\mathbf{M}_z$ was multiplied by the convolution matrix $\mathbf{M}_\mathrm{water}$ of the water signal to get a $\delta$ - like pulse for the pure water-signal without fatty tissue in the measurement chamber.} 
\label{Fig:Fat_tissue_1D}
\end{figure}

For excitation of plane acoustic waves the surface of a silicon wafer was illuminated by a nanosecond laser pulse (Fig.~\ref{Fig:Fig_Setup_plane_waves}). Due to the abrupt local heating due to absorption of the 532 nm wavelength laser light pulse, the silicon wafer expands thermoelastically and generates an acoustic wave, propagating through a layer of fatty tissue and is detected by an unfocused ultrasound transducer (V358-SU, Panametrics, Waltham, MA). The measured acoustic pressure without fat, through 6 mm thick porcine fat tissue, and through 20 mm thick tissue is shown in Fig. \ref{Fig:Fat_tissue_1D}(a) as a function of time \cite{Burgholzer.2020,Burgholzer.2020b}. For comparison, the measured signals are scaled and time-shifted. The frequency spectrum was calculated by Fourier transformation in time and the frequency dependent attenuation for the 6 mm and 20 mm fat layer was determined by dividing their spectrum through the spectrum of the reference signal without fat. It turned out that a power law $\alpha(\omega)=\alpha_0 \omega^n$ with an exponent $n=1.5$ and $\alpha_0=0.87 \text{dB} \, \text{MHz}^{-n}\text{cm}^{-1}$ fits the attenuation in a wide frequency range very well \cite{Burgholzer.2020}. 

The experimental set-up now is slightly different compared to the photoacoustic imaging set-up described in section \ref{sec:Imaging_model}. In photoacoustic imaging, the short excitation pulse at the time zero excites a pressure wave in the whole sample at the same time. If an acoustic signal arrives later at the detector this comes from a longer propagation distance and causes a higher attenuation. Here, the fat layer always has the same thickness and therefore the attenuation for earlier or later arriving parts of the signal at the detector is the same. Only for the excitation as a Dirac delta in space and time as described in Eq. (\ref{Eq:attenuated_pressure_wave}), this is the same. Since the differential equations are linear, the general attenuated signal is a temporal convolution of the input signal with the attenuated solution for Dirac excitation, or a simple multiplication in the frequency domain\cite{DeanBen.2011}. The discrete version of the relationship between the measured pressure and the virtual pressure wave now, instead of Eq. (\ref{Eq:Matrix_pressure}), becomes 
\begin{equation}
    \mathbf{p} = \mathbf{M}_z \mathbf{p}_\text{virt} \,,
    \label{Eq:Matrix_pressure_convolution}
\end{equation}
\noindent
where $\mathbf{p}$ and $\mathbf{p}_\text{virt}$ are the vectors of the attenuated and virtual pressure signal at discrete time steps, respectively. Writing the discrete Fourier and inverse Fourier transformation as multiplication by $\mathbf{F}$ and its conjugate transpose $\mathbf{F}^*$, and  $\operatorname{diag}(.)$ forms a diagonal matrix, $\mathbf{M}_z$ can be written by neglecting the dispersion as \cite{Burgholzer.2020}
\begin{equation}
    \mathbf{M}_z  = \mathbf{F}^*\operatorname{diag}(\exp{(-\alpha_0 \omega^n z)})\mathbf{F},
    \label{Eq:M_z}
\end{equation}
\noindent
which immediately shows that the singular values of $\mathbf{M}_z$ decay exponentially and according to Eq. (\ref{Eq:Photoacoustic_cut_off}) for larger frequencies than the truncation frequency $\omega_\text{cut}$ the amplitude of these wave components is damped below the noise level, which results in

\begin{equation}
    \text{SNR} \exp{(-\alpha_0 \omega_\text{cut}^n z)} = 1 \phantom{X} \text{or} \phantom{X} \omega_\text{cut} = \sqrt[n]{\frac{\ln(\text{SNR})}{\alpha_0z}}.
    \label{Eq:Convolution_cut_off}
\end{equation}

The spatial resolution according to Eq. (\ref{Eq:delta_resolution}) is half the wavelength at the truncation frequency $\omega_\text{cut}$. For the SNR of 1358 (63 dB), the truncation frequency for a fat thickness of 6 mm and 20 mm is 24 MHz and 11 MHz, respectively, which corresponds to a spatial resolution of 32 $\mu$m after 6 mm of fat and 70 $\mu$m after 20 mm of fat. \par
The acoustic attenuation in water compared to fat can be neglected \cite{Burgholzer.2020}. Therefore, the signal measured in water without fat can be taken as the virtual pressure signal $\mathbf{p}_\text{virt}$, which is no single Dirac delta pulse, but gets negative and shows additional "ringing" because of the laser ultrasound excitation within the silicon wafer and the characteristic of the piezoelectric transducer and the amplifier. All these influences on the signal can be described by a matrix $\mathbf{M}_\mathrm{water}$ with $\mathbf{p}_\text{virt} = \mathbf{M}_\mathrm{water} \mathbf{p}_\delta$, and Eq. (\ref{Eq:Matrix_pressure_convolution}) can be written as 
\begin{equation}
    \mathbf{p} = \mathbf{M}_z \mathbf{M}_\mathrm{water} \mathbf{p}_\delta.
    \label{Eq:Matrix_pressure_convolution_delta}
\end{equation}
\noindent
Using the truncated SVD for the inversion of $\mathbf{M}_z \mathbf{M}_\mathrm{water}$, the reconstructions of $\mathbf{p}_\delta$ from the measurements $\mathbf{p}$ for water, 6 mm fat, and 20 mm fat are shown in Fig. \ref{Fig:Fat_tissue_1D}(b). Multiplying the temporal width of these peaks of 21 ns and 46 ns by the sound velocity of 1512 m/s in fatty tissue perfectly fits to the resolution limits of 32 $\mu$m after 6 mm of fat and 70 $\mu$m after 20 mm of fat derived above from the truncation frequencies.\par

\subsubsection{Axial thermal profiling of planar heat sources}
\label{sec:1D_depthprofiling_PT}

\begin{figure}
\includegraphics[trim={8.5cm 3.5cm 8.5cm 3.5cm},clip,width=\linewidth]{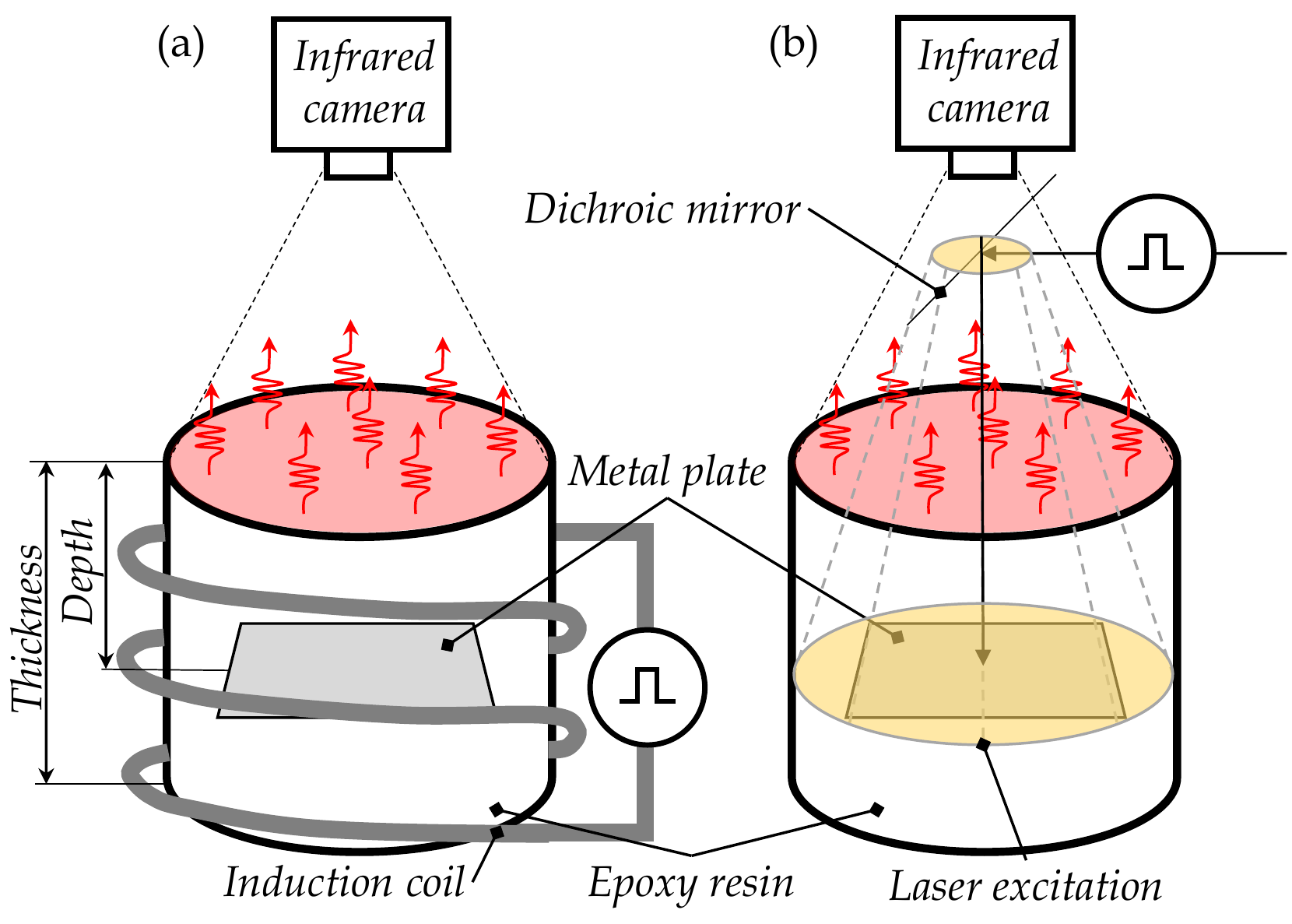}
\caption{\label{fig:Fig_temperature_depth2} Experimental setup for 1D thermographic reconstruction of internal heat sources: a thin foil (thickness approx. 0.2 mm) embedded in epoxy resin is heated inductively (a) or by the absorption of laser light (b). Since the sample diameter (40 mm) is significantly larger than the deepest defect depth (11.7 mm), we can assume one-dimensional heat conduction in the axis of symmetry.} 
\label{Fig:Fig_PT_1D_experiment}
\end{figure}

To generate an internal planar heat source for an experimental study, a thin plate (foil) was embedded in pure epoxy resin. Laser light is absorbed for thermal excitation and heats the foil (Fig.\ref{Fig:Fig_PT_1D_experiment} (b)). In the case of an electrically conductive film, electromagnetic induction can also be used to heat the film  (Fig.\ref{Fig:Fig_PT_1D_experiment} (a)). The epoxy resin is opaque in the spectral sensitivity range of the mid-wave IR camera (3 to 5 $\mu$m) . Therefore, the measured infrared radiation only comes from the sample surface (z = 0 mm).

\begin{figure}
\includegraphics[width=\columnwidth]{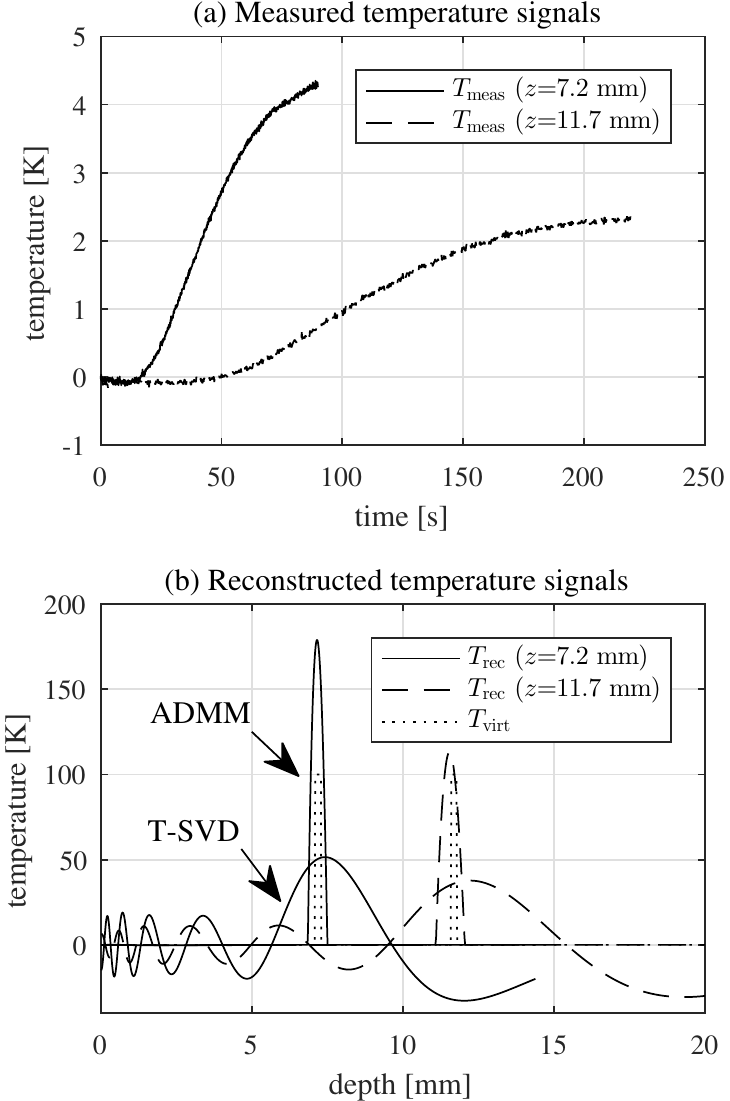}
\caption{Depth resolved thermal source profiling: (a) Thermal transients resulting from inductive pulse heating for two different layer sources with a layer thickness of 0.2 mm at a depth $z$ of 7.2 mm and 11.7 mm and (b) the solution of regularized inverse problem using the direct method of T-SVD and the iterative procedure ADMM including prior information. The reconstructed depth profiles $T_\mathrm{rec}$ are compared with the ideal source distribution $T_\mathrm{virt}$.} 
\label{Fig:Fig_PT_1D_imaging}
\end{figure}

 The thermal diffusivity of the epoxy sample was obtained by the Parker method\cite{Parker.1961}  as $\alpha = 1.4 \times 10^{-7} \mathrm{m}^2 \mathrm{s}^{-1}$. For two different epoxy samples, with the embedded film at a depth of $z$ = 7.2 mm and 11.7 mm, the surface temperature as a function of time has been measured with a quantum IR detector (Fig. \ref{Fig:Fig_PT_1D_imaging}(a)). Due to the low temperature increase on the sample surface, the measurement noise is approximately additive white Gaussian noise with a standard deviation of 25 mK\cite{Breitwieser.2020}. The measurement parameters are listed in Tab. \ref{tab:measurement_parameter_1D}, with the pulse duration $t_\mathrm{p}$, the number of discretizations in time and space $N$ and the temporal and spatial discretization $\Delta t$ and $\Delta z$. 

\begin{table}
\caption{\label{tab:measurement_parameter_1D} Measurement and reconstruction parameters for the 1D reconstruction of the internal heat sources. The SNR was derived from the considered singular values.}
\begin{ruledtabular}
\begin{tabular}{ccccc|ccc}
$z$ & $t_\mathrm{p}$ & N & $\Delta t$ & $\Delta z$ & SNR & $\omega_\mathrm{cut}/(2\pi)$ & $\delta_\mathrm{resolution}$ \\
$[\text{mm}]$ & $[\text{s}]$ & $[-]$ & $[\text{s}]$ & $[{\mu\text{m}}]$ & $[-]$ & $[\text{mHz}]$ & $[\mathrm{mm}]$\\
\hline
7.2 & 0.5 & 900 & 0.1 & 17 & 587 & 36 & 3.5 \\
11.7 & 1  & 1100 & 0.2 & 24 & 930 & 16 & 5.4 \\
\end{tabular}
\end{ruledtabular}
\end{table}

We invert these experimental temperature signals into virtual wave signals to reconstruct the location, the width and amplitude of the internal heat sources, showing the capability of incorporating prior information in the regularization. The reconstructed depth profile of the initial temperature using the direct regularization method T-SVD \cite{Hansen.1998} is shown in Fig. \ref{Fig:Fig_PT_1D_imaging}(b). The comparison of the different reconstructed depth profiles clearly shows the entropy production caused by thermal diffusion that is equal to information loss with the result of lower axial resolution for the deeper defect, as listed in Tab. \ref{tab:measurement_parameter_1D}. The SNR, which was derived from the singular values, is nearly doubled for the defect depth of 11.7 mm, as the duration $t_p$ of the heating pulse is increased from 0.5s to 1s. The truncation frequency $\omega_\mathrm{cut}$ and the spatial resolution $\delta_\mathrm{resolution}$ are derived by Eq. (\ref{Eq:kcut}) and Eq. (\ref{Eq:delta_resolution_thermal}), respectively.\par
Using additional information about the system e.g. through regularization constraints improves the definition of the system, decreases the available states and leads to lower entropy, which leads to an increased spatial resolution in image reconstruction. To increase the information content in the regularization process, we introduce prior knowledge in form of positivity, which is reasonable due to the solely non-negative amplitude values of the virtual wave, as well as for the photoacoustic pressure, in the 1D regime \cite{Wang.2009}. In contrast to the bipolar reconstruction with T-SVD, non-negative values for the reconstructed virtual wave field can be enforced using iterative regularization procedures, such as the Alternating Direction Method of Multipliers (ADMM)\cite{Parikh.2014,Boyd.2010}. Additionally, for a majority of practical problems in photothermal imaging, we can assume sparsity, since in the application of defect detection mostly isolated singular defects occur, e.g. cracks, pores, delaminations or inclusions of other materials, or in the application of parameter estimation, only the positions of the sources, the back wall or individual inner boundary layers should be identified. Consequently, we have only a few point or line scatterers, which leads to a sparse virtual wave signal. Sparsity is introduced in the ADMM by an appropriate formulation of the objective function using $\ell^1$-norm minimization. As shown in the reconstructed depth profiles in Fig. \ref{Fig:Fig_PT_1D_imaging}(b), iterative regularization with prior information leads to an improvement in energy localization and hence to a higher axial resolution. 

\subsection{Active thermographic computed tomography (ATCT)}
\label{sec:ATCT}

Active thermographic computed tomography (ATCT) is a new hybrid reconstruction technique that utilizes the photothermal (PT) effect\cite{Thummerer.2020,Thummerer.2020b}, or other thermal excitation sources as inductive heating\cite{Burgholzer.2017,Burgholzer.2018}, mechanical friction through ultrasound, microwaves, for signal generation. In the case of photothermal computed tomography (PTCT) an optical pulse is used to irradiate an object, such as biological tissue or manufactured materials and products, to generate thermal waves within the object. The thermal waves diffuse to the sample surface, where the temperature change can be measured using infrared cameras. Due to the focal plane array of the infrared camera  up to $10^6$ signals can be recorded simultaneously, which is equivalent to an extremely large aperture in contrast to photoacoustic transducers. The aim of PTCT is to reconstruct an image that represents a map of the initial temperature distribution within the object from the measured photothermally-induced thermal signals. The initial temperature distribution is proportional to the absorbed optical energy, which can reveal useful information of the internal structure, such as material defects, or inhomogeneous material parameters. \par
The reconstruction process in ATCT is a two step algorithm:
\begin{itemize}
    \item \textit{Computation of the virtual wave field}: The original thermographic problem is converted to a photoacoustic imaging task by a pointwise transformation, that means separately for each pixel, of the measured temperature field into a virtual wave field as shown in section (Sec.\ref{sec:1D_depthprofiling_PT}). 
    \item \textit{Virtual wave back-propagation}: The resulting virtual wave field exhibits wave properties, such as wavefront propagation, reflection and refraction. Frequency domain-synthetic aperture focusing technique (F-SAFT), a well known acoustic reconstruction method \cite{Levesque.2002,Busse.1992}, can be applied to image the initial temperature distribution. This allows a multidimensional image reconstruction with the advantage of a higher SNR.
\end{itemize}

The common inverse heat conduction solutions for multidimensional thermographic imaging leads to large scale optimization problems with high computation costs. The virtual wave concept for ATCT splits the severely ill-posed 3D large scale problem into a variety number of small-scale 1D problems, where efficient algorithms, e.g. ADMM, can compute parallel for the solution. The 1D depth-encoded signals calculated for each location could be gathered and then reconstructed to a 3D image using F-SAFT with fast computational architecture, which could be also accelerated by graphical processing unit (GPU) programming \cite{Liu.2018}. \par
In 3D imaging, we prefer the regularization technique ADMM including the prior information sparsity and positivity because of its improvements in terms of sensitivity and depth resolving capability compared to direct regulizers. The energy input during thermal excitation always results in positive temperature signals. By converting the temperature field into a virtual wave field, positivity cannot be applied directly for 2D
or 3D wave propagation. An initially non-negative acoustic signal will take negative values during propagation \cite{Wang.2009}. The circular and spherical projections,which is in 2D the Abel transformation and in 3D the time integral of the virtual wave \cite{Burgholzer.2007b}, retain positivity of the initial source. For one data point the information gain by a positivity constraint would be only a factor of 2, but for a signal with $N$
data points this factor becomes $2^N$, which can be large for higher $N$.\par
In the next sections we show two examples of ATCT with limited data (limited view), when the detectors cannot be placed around the complete object. In practical measurements, only a limited space around the specimen is available for photothermal detection. The thermal diffusivity and the resulting virtual speed of sound of the bulk is assumed to be constant and isotropic.    

\subsubsection{Reconstruction from a single detector plane}
\label{sec:single_detector_plane}

\begin{figure}
\includegraphics[width=0.9\columnwidth]{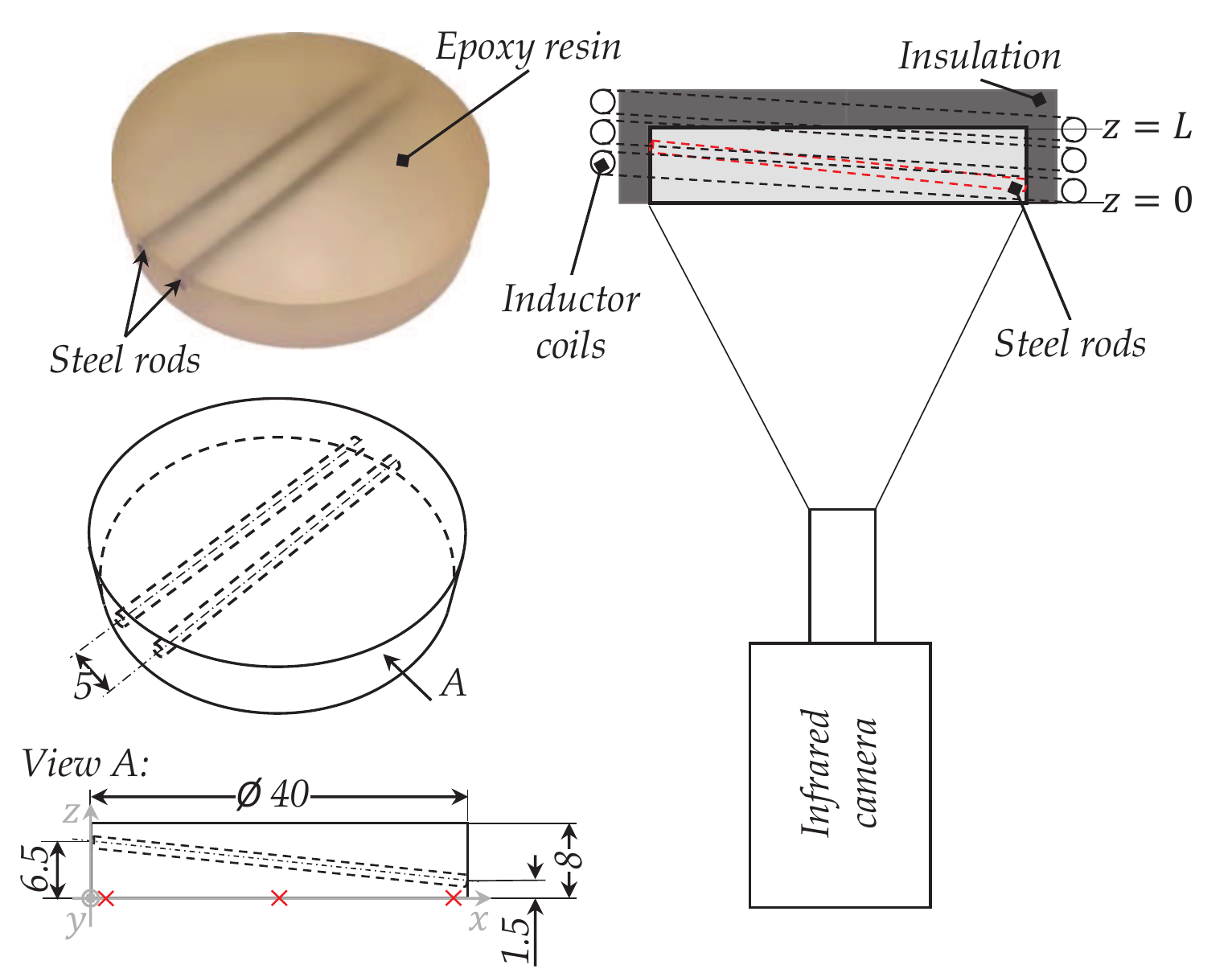}
\caption{Experimental setup: (a) Test specimen built up with steel rods with a diameter of 1.5 mm, that are embedded
in epoxy resin and in (b) the principle sketch of the measurement set-up: The
steel rods are stimulated by induction of eddy current. The resulting change of the surface temperature is measured with an infrared camera.} 
\label{Fig:Fig_TestSpecimen_2D}
\end{figure}

The experimental phantom is built up with two steel rods, which are embedded in epoxy resin (Fig. \ref{Fig:Fig_TestSpecimen_2D}). The cylindrical axes of the steel rods are parallel with a distance of 5 mm and inclined to the object surface. The inductive heated steel rods work as volumetric heat sources with a heating time of $t_\mathrm{p}$ = 2 s similar to the 1D photothermal example (Sec. \ref{sec:1D_depthprofiling_PT}). The temperature field is measured on a single surface at z=0. The measured signal decreases with increasing depth of the rods, due to the homogeneous diffusion in each direction. The spatial- and time discretizations were $\Delta_x = 168 \, \mu m$ and $\Delta_t = 40 \, \text{ms}$. \par
\begin{figure}
\includegraphics[width=1.0\columnwidth]{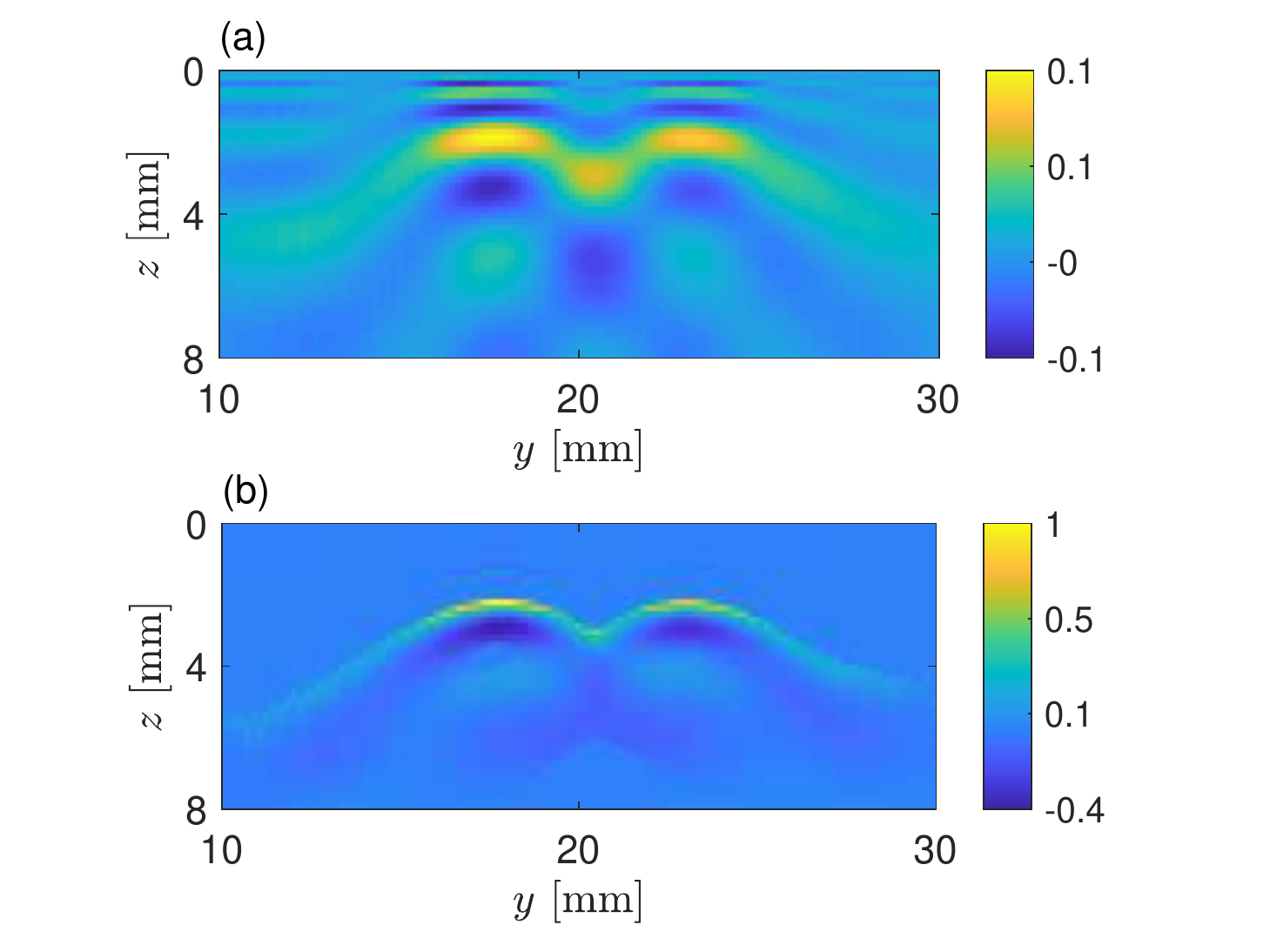}
\caption{2D cross section of the reconstructed reconstructed virtual wave field of the two steel bars at $x$ = 3 mm produced by use of two different regularization techniques, where (a) the direct method of T-SVD and (b) the iterative procedure of ADMM with prior information is used. The virtual scattering hyperbolas are visible.} 
\label{Fig:Fig_PT_2D_VW_field}
\end{figure}

Reconstruction results based on this phantom and T-SVD regularization were also published by Burgholzer et al. \cite{Burgholzer.2018}. In this tutorial, not only T-SVD is used for regularization, but also the iterative regularization scheme ADMM is used, which enables the incorporation of prior information in the form of positivity and sparsity for defect reconstruction. A cross section of the virtual wave field reconstructed by T-SVD and ADMM is illustrated in Fig. \ref{Fig:Fig_PT_2D_VW_field}. The virtual wave fields were normalized to the maximum wave amplitude, that was obtained by ADMM regularization. Each cross section shows the typical scattering hyperbola as consequence of the cylindrical steel rods, but the ADMM regularization leads to a much sharper localization. Based on the virtual wave field, the initial temperature distribution can be calculated applying inverse wave propagation methods, like F-SAFT.

\begin{figure*}
\includegraphics[trim={3cm 0cm 3cm 0cm},clip,width=\linewidth]{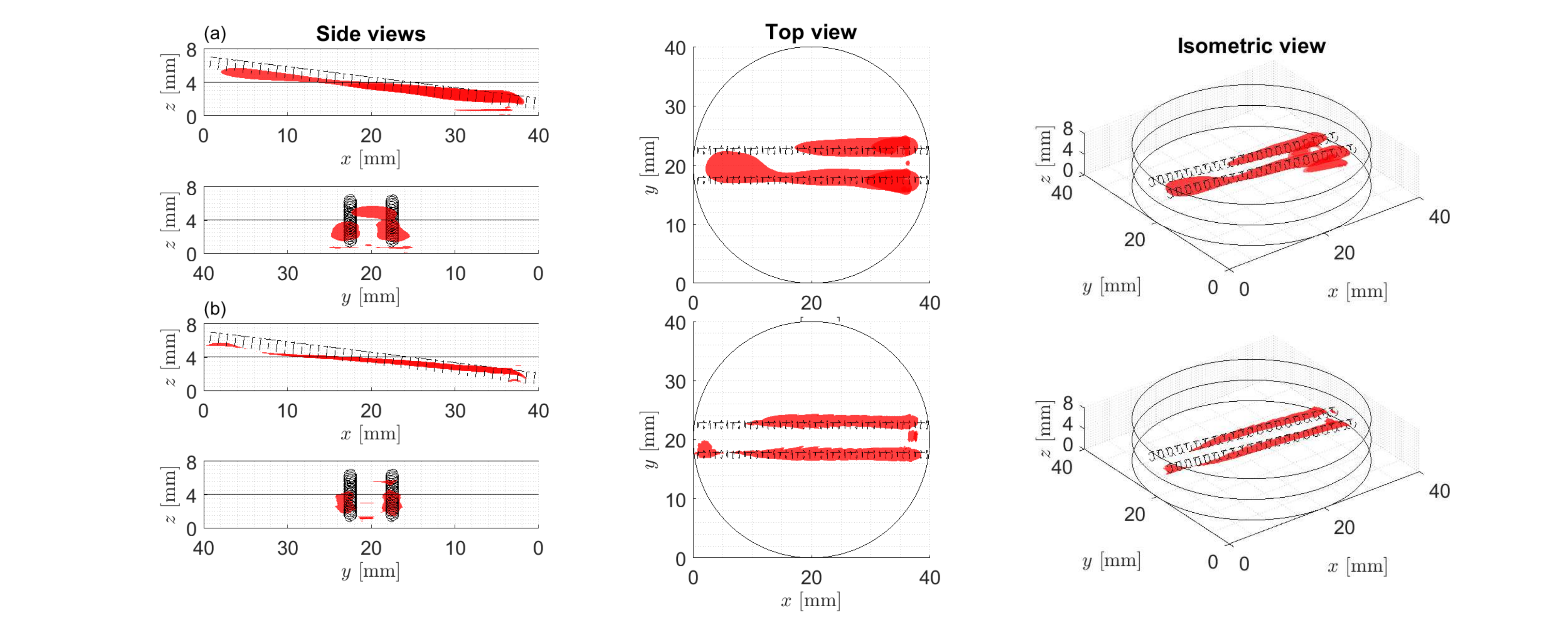}
\caption{Isosurface illustration of the reconstructed internal heat sources obtained via a) T-SVD and b) ADMM employing F-SAFT.}
\label{Fig:View3DSteelRodsInEpoxyRed}
\end{figure*}

Fig. \ref{Fig:View3DSteelRodsInEpoxyRed} illustrates the resulting 3D initial temperature distribution for the different regularization methods. The amplitude of the reconstructed initial temperature field was binarized with a threshold value to represent the localization of thermal sources at a point within the 3D sample volume. The circles indicate the real position of the steel rods in the epoxy cylinder. Fig. \ref{Fig:View3DSteelRodsInEpoxyRed} (a) exhibits the T-SVD based reconstruction. It is visible, that the steel rods can be distinguished till $z$ = 4 mm, as expected for direct reconstructions without any prior information. The resolution is in the range of the depth of the sources. Moreover, artefacts are present in regions with low SNR. Fig. \ref{Fig:View3DSteelRodsInEpoxyRed} (b) illustrates the reconstruction employing ADMM. In this process, positivity and sparsity was introduced to improve the regularized solution.  Here, the steel rods are separable in the domain 10 < x < 40 mm. Moreover, the artefacts are significantly reduced compared to T-SVD.

\subsubsection{3D plane detection}

\begin{figure*}
\includegraphics[trim={0cm 2cm 0cm 1.5cm},clip,width=\linewidth]{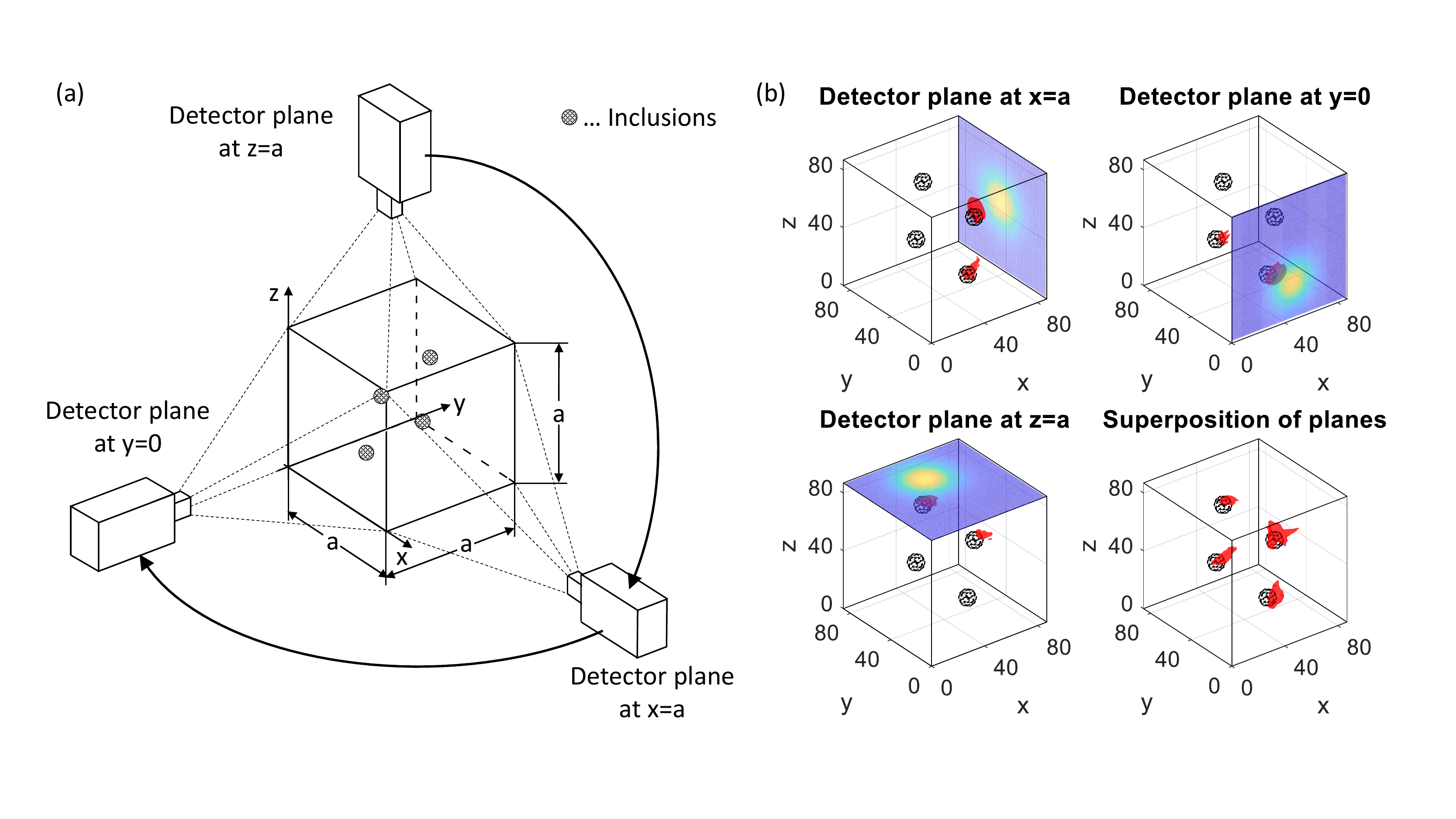}
\caption{(a) Experimental setup and detection planes of the ATCT measurements and (b) the isosurface illustration of the reconstructed internal heat sources obtained with ADMM for three different detector planes and a superposition of the single detector plane reconstructions. The steel spheres had depths of 4.3 mm, 7.5 mm and 10.7 mm from the detection planes.}
\label{fig:ThermographicTomographie_PrincipalSketch}
\end{figure*}

An example of ATCT with temperature data from three orthogonal detection planes is shown in Fig. \ref{fig:ThermographicTomographie_PrincipalSketch}. The phantom is an epoxy cube with an edge length of $a$ = 15 mm containing four spherical steel spheres with a diameter of 2 mm. The epoxy material, the steel beads, the inductive heating device and the position of the sample inside the coil were the same as in Sec. \ref{sec:single_detector_plane}. The spatial and temporal discretization of the IR camera were $\Delta x = \Delta y = \Delta z = 172 \, \mu$m and $\Delta t$ = 50 ms. The bottom and the sidewalls of the cube were thermally isolated to ensure adiabatic boundary conditions. The time-dependent temperature fields were measured on the right side plane ($x = a$), front plane ($y = 0$) and on the top plane ($z = a$) for a time duration of 120 s after the pulse excitation. The pulse time of 7.5 s was chosen to obtain also thermographic signals from the deeper spheres. \par 
Reconstructions based on this phantom and T-SVD regularization were also published in Burgholzer et al. \cite{Burgholzer.2017}. In this tutorial and in contrast to prior works, we use ADMM regularization and incorporate prior information in the form of positivity and sparsity to improve the quality of the regularized solution. The three corresponding reconstructions using ADMM for regularization and F-SAFT for image reconstruction are shown in Fig. \ref{fig:ThermographicTomographie_PrincipalSketch}. For all three reconstruction planes, limited view artifacts are visible and the inclusions far from the detector planes cannot be reconstructed at all.        

\section{Summary, Conclusions and Outlook}
\label{sec:Summary}
In nature, real processes are always irreversible and show a clear direction of the arrow of time. Some processes, such as the propagation of an acoustic wave, especially for low frequencies in a medium with low viscosity, are in a good approximation reversible. The corresponding wave equation for pressure is symmetric in time; for every pressure wave, which is a solution of this equation, the time-reversed pressure wave is also a solution. Even if some acoustic waves can be described in a good approximation by the wave equation, acoustic attenuation cannot be avoided totally. Therefore, we call this ideal solution a "virtual" wave, in comparison to the real measured signal. In this tutorial it is demonstrated for thermal and acoustic waves, that the measured and the virtual signal are linked by a local relation (Eq. \ref{Eq:linear_relation}).\par
For non-destructive imaging of subsurface structures, from an information theoretical viewpoint it is advantageous to use "nearly reversible" waves to transport the information from the structures to the sample surface, such as acoustic waves with low attenuation. Due to heat diffusion thermal waves exhibit a higher entropy production, which is shown to be equal to the information loss (section \ref{sec:thermodynamic}) and thus results in a lower spatial resolution (Tab. \ref{tab:truncation_frequency} and \ref{tab:truncation_frequency2} or Fig. \ref{fig:Fig_pressure_depth_1mm} and \ref{fig:Fig_temperature_depth}). In section \ref{sec:thermodynamic} two simple examples for stochastic processes are given, where the close connection between fluctuation and dissipation ("kicked Brownian particle") and diffusion ("ideal gas diffusion between two boxes") can be shown explicitly.\par
In section \ref{sec:imaging} the information theoretical cut-off criterion using the Chernoff-Stein's Lemma (section \ref{sec:thermodynamic}) is applied in frequency domain to photoacoustic and photothermal imaging. The information related criterion gives the same truncation value as when the signal amplitude gets less than the noise level. This is always the case if the mean energy of the wave component and therefore also its information content is proportional to the square of the signal amplitude. In section \ref{sec:1Dimaging} for the 1D imaging even in fat with rather high acoustic damping the resolution is always much better than for thermographic imaging (see also the simulation in Glycerine in section \ref{sec:compensation_acoustic_attenuation}). Therefore, for thermographic imaging with this high entropy production from heat diffusion, it is essential to use additional information such as positivity and sparsity by implementing iterative algorithms, e.g. by ADMM to get a better resolution, also in the sub-mm range (Fig. \ref{Fig:Fig_PT_1D_imaging}(b)). This has been used in thermographic computed tomography (section \ref{sec:ATCT}), either by 2D reconstruction from single detector plane measurements or on three perpendicular planes for a cubic sample containing steel spheres to be imaged.\par
Using the T-SVD without incorporating additional information for a single 1D reconstruction, the spatial resolution does not get better compared to a direct inversion of the heat diffusion. The advantage of the virtual wave concept in 1 D is, that more advanced regularization techniques, incorporating a priori information such as sparsity or positivity can be utilized in the reconstruction process. Including additional prior information allows thermographic reconstruction with a significantly better resolution than without that additional information, as shown in Fig. \ref{Fig:Fig_PT_1D_imaging}(b), Fig. \ref{Fig:Fig_PT_2D_VW_field}, and Fig. \ref{Fig:View3DSteelRodsInEpoxyRed}\cite{Thummerer.2020,Thummerer.2020b}. In 2D and 3D, after the computation of the virtual wave field, well known acoustic reconstruction methods, such as F-SAFT are used (section \ref{sec:ATCT}). Here, the SNR is significantly enhanced as the heat flow in all directions is taken into account for reconstructions. In the past, one-dimensional axial image reconstruction was mainly carried out for thermographic reconstruction regardless of the lateral heat flow. The virtual wave concept respects the lateral heat flow, which yields an improved SNR and therefore an increased spatial resolution.\cite{Burgholzer.2017}. This is similar to averaging of many 1D measurements when imaging a layered structure, but with the essential advantage, that the virtual wave concept can be used for any 2D or 3D structure to be imaged.\par
For future work, deep learning neural networks are planned to be used as the second step instead of e.g. F-SAFT. For the same date from Fig. \ref{Fig:View3DSteelRodsInEpoxyRed} this gives another significant enhancement in resolution. In Fig. \ref{Fig:View3DSteelRodsInEpoxyRed} the threshold-level for the isosurface plot is rather critical to avoid additional artifacts. With a deep neural network trained only by simulated data the reconstructions are very stable and choosing a threshold-level for the isosurface plot is not critical as reconstruction artifacts hardly appear \cite{Kovacs.542020582020}.

\begin{acknowledgments}
The financial support by the Austrian Federal Ministry of Science, Research and Economy and the National Foundation for Research, Technology and Development is gratefully acknowledged. Furthermore, this work has been supported by the project “multimodal and in-situ characterization of inhomogeneous materials” (MiCi), by the the federal government of Upper Austria and the European Regional Development Fund (EFRE) in the framework of the EU-program IWB2020. Financial support was also provided by the Austrian research funding association (FFG) under the scope of the COMET programme within the research project “Photonic Sensing for Smarter Processes (PSSP)” (contract number 871974). This programme is promoted by BMK, BMDW, the federal state of Upper Austria and the federal state of Styria, represented by SFG. Parts of this work have been supported by the Austrian Science
Fund (FWF), projects P 30747-N32 and P 33019-N.\par
\noindent
The data that support the findings of this study are available from the corresponding author upon reasonable request.
\end{acknowledgments}

\end{document}